\documentclass[12pt]{article}

\usepackage{amssymb,latexsym,amsmath}

\usepackage{fullpage}
\usepackage{nicefrac}
\usepackage{mathrsfs}%different \cal using \mathscr
\usepackage{slashed}% for dslash

\usepackage[pdftex]{graphicx}

\usepackage{bbm}

\makeatletter \@addtoreset{equation}{section}
\makeatother

\newcommand{\be}{\begin{equation}}
\newcommand{\ee}{\end{equation}}
\newcommand{\bea}{\begin{eqnarray}}
\newcommand{\eea}{\end{eqnarray}}
\newcommand{\nn}{\nonumber}
\newcommand{\re}{\mathrm{Re}}
\newcommand{\im}{\mathrm{Im}}
\newcommand{\vep}{\varepsilon}

\newcommand\mathbbone{\ensuremath{\mathbbm{1}}}

\begin{document}

\begin{titlepage}
	\thispagestyle{empty}
	\begin{flushright}
		\hfill{DFPD-11/TH/1}
	\end{flushright}
	
	\vspace{90pt}
	
	\begin{center}
	    { \LARGE{\bf Constructing Lifshitz solutions from AdS}}\\ [3mm]

			\vspace{30pt}
		
		{Davide Cassani$^{\,1}$ and Anton F. Faedo$^{\,2}$}
		
		\vspace{25pt}
		
		{\small
		{\it ${}^1$  Dipartimento di Fisica ``Galileo Galilei''\\
		Universit\`a di Padova, Via Marzolo 8, 35131 Padova, Italy}
		
		\vspace{15pt}
		
		{\it ${}^2$  INFN, Sezione di Padova \\
		Via Marzolo 8, 35131 Padova, Italy}
		
		\vspace{15pt}

		{cassani, faedo @ pd.infn.it}
		}
		
		\vspace{90pt}

		{\bf Abstract}
	\end{center}
		
Under general assumptions, we show that a gravitational theory in $d+1$ dimensions admitting an AdS solution can be reduced to a $d$-dimensional theory containing a Lifshitz solution with dynamical exponent $z=2$. Working in a $d=4$, $\mathcal N=2$ supergravity setup, we prove that if the AdS background is $\mathcal N=2$ supersymmetric, then the Lifshitz geometry preserves 1/4 of the supercharges, and we construct the corresponding Killing spinors.
We illustrate these results in examples from supersymmetric consistent truncations of type IIB supergravity, enhancing the class of known 4-dimensional Lifshitz solutions of string theory. As a byproduct, we find a new AdS$_4\times S^1\times T^{1,1}$ solution of type IIB.

\vspace{10pt}

\end{titlepage}

\baselineskip 6 mm

%%%%%%%%%%%%%%%%%%%%%%%%%%%%%%%%%%%%%%%%%%%%%%%%%%%%%%%%%%%%%%

%%%%%%%%%%%%%%%%%%%%%%%%%%%%%%%%%%%%%%%%%%%%%%%%%%%%%%%%%%%%%%
\section{Introduction}

The gauge/gravity correspondence is a fascinating theoretical achievement and a precious tool for studying the strong coupling dynamics of gauge theories.
In the past years most of the developments were directed to model QCD-like theories holographically, and new insight into phenomena like confinement and the quark-gluon plasma has been gained.
Only very recently the scope of the correspondence has been broadened to encompass other setups, mostly coming from solid state physics. While in particle physics there is mainly one gauge theory one wants to dualize, that is QCD, in condensed matter physics a large number of different systems is studied, and many can be strongly coupled. It is tempting then to try to develop the corresponding gravity duals. 
However, their properties are frequently far from our high energy intuition. For example, Lorentz invariance is not such a fundamental symmetry. This shows that, in order to model these systems holographically, one often needs to depart from the usual AdS solutions in a quite unconventional way. 

In this context, an instance that has attracted considerable attention is that of phase transitions dominated by fixed points with a scaling symmetry of the anisotropic form
\be\label{scaling}
t\,\rightarrow\, \lambda^z\,t\,,\qquad\qquad\qquad\qquad x^\ell\,\rightarrow\lambda\, x^\ell\,,
\ee
where $t$ and $x^{\ell}$ are the temporal and spatial directions, respectively, while $z$ is called the dynamical exponent.
The particular value $z=1$ corresponds to the scale invariance of the relativistic conformal group. In \cite{Kachru}, a gravity dual capturing this symmetry was constructed in $d$ dimensions as a generalization of the AdS geometry to the family of Lifshitz metrics
\begin{equation}\label{LifMetric}
ds^2(M_d) \,=\, L^2\Big[  -r^{2z}\,dt^2+r^2\,dx^\ell dx^\ell +\frac{dr^2}{r^2} \,\Big]\,,\qquad \ell=1,\ldots,d-2\,,
\end{equation}
which are invariant under (\ref{scaling}) provided one scales the holographic coordinate as $r\,\rightarrow\, \lambda^{-1}\,r$. In the original proposal these metrics arose as solutions to a 4-dimensional model involving a vector and a 2-form coupled via a topological term. Later it was realised that the same metrics, in arbitrary dimensions, can be supported by a timelike massive vector \cite{Taylor}. 

An important point is that the approach of \cite{Kachru, Taylor} was phenomenological, that is, the models under study were {\it ad hoc} constructions without any relation to string theory. Consequently, it is not clear if the solutions found in this way provide legitimate duals. 

The quest for embedding Lifshitz geometries into string theory started with some negative statements and even no-go theorems \cite{NoGo, Blaback}. Some partial results appeared in \cite{Azeyanagi, Nameki, Hartnoll}, while the first proper solution was constructed in \cite{BalNar}, followed by  \cite{DonosGauntlett_Lif}, \cite{Gregory} and \cite{Donos39}.
Whilst these solutions can surely be found working directly at the 10- or 11-dimensional level, it is often convenient to look for them in more restricted lower-dimensional setups. The proper way to ensure an embedding into string theory is to work with consistent truncations. These have proven very useful in finding string theoretic solutions dual to condensed matter systems, like superconductors \cite{Gubser_sup, Gauntlett_sup} and models with non-relativistic Schr\"odinger invariance \cite{Maldacena}. In this context, the complications in finding Lifshitz solutions to string theory can be traced back to the difficulties in finding consistent truncations with suitable massive vectors to support them, as already stated in \cite{Taylor}. 
Nowadays we have access to various consistent truncations of 10- and 11-dimensional supergravity containing massive modes \cite{Maldacena, Gauntlett_11, ExploitingN=2, IIBonSE, Gauntlett_SE, Liu, Skenderis:2010vz, T11, Bena, Donos39}, but none of them seemed to include a vector with the correct mass (except the one in \cite{Donos39}). A key point in \cite{BalNar, DonosGauntlett_Lif} was to identify the appropriate vector, coming from gauging the reparameterizations of a circle coordinate.

As noticed in \cite{DonosGauntlett_Lif}, the Lifshitz solutions with dynamical exponent $z=2$ found in \cite{BalNar} can be seen as a circle reduction of certain geometries with non-relativistic scaling symmetry of the Schr\"odinger type \cite{Son, BalaMcGreevy} with $z=0$, which in turn are simple deformations of AdS space. 
These solutions can all be obtained from a model with a free massless scalar and a negative cosmological constant \cite{BalNar, Costa}. However, consistent truncations typically come with a non-trivial scalar potential instead of a cosmological constant, which makes the embedding into string theory harder.

In this paper, we provide a general framework in which this problem can be solved by relating Lifshitz solutions in $d$ dimensions to AdS solutions in $d+1$ dimensions. More specifically, we prove that a $(d+1)$-dimensional gravity theory with general matter couplings and admitting an AdS vacuum can be reduced on a circle $S^1$ to provide a theory admitting a Lifshitz$_d$ solution with $z=2$. This is located precisely at the values of the scalars that extremize the $(d+1)$-dimensional potential. In order for the proof to hold, we need the existence of an axion in the $(d+1)$-dimensional theory. This allows to introduce a flux term along $S^1$, which gives mass to the vector gauging the circle isometry. The value of the mass turns out to be the adequate one to support the Lifshitz metric.

When the model comes from a consistent truncation, we enhance the class of Lifshitz solutions to string theory, because the metric on the compact manifold transverse to the $(d+1)$-dimensional spacetime is not required to be Einstein, contrary to~\cite{DonosGauntlett_Lif}. 

We also study the supersymmetry conditions for Lifshitz solutions. These are expected to provide a better controlled arena in the perspective of understanding the gauge-gravity dictionary. Working in a $d=4$, $\mathcal N=2$ gauged supergravity setup with general matter couplings, we find that if the starting AdS$_5$ vacuum preserves $\mathcal N=2$ supersymmetry, then the associated Lifshitz$_4$ background inherits 1/4 of the supercharges. Besides, we construct the corresponding Killing spinors, and we discuss the relation between the supersymmetry conditions and the equations of motion. 

In the second part of the paper, we implement our general results by considering some 5-dimensional $\mathcal N=2$ supersymmetric consistent truncations of type IIB supergravity, elaborated in \cite{IIBonSE, T11} and based on squashed Sasaki--Einstein manifolds. On the one hand, these models yield new Lifshitz solutions in correspondence to certain non-Einstein metrics supporting AdS$_5$ vacua \cite{RomansIIBsols}.
On the other hand, they provide a concrete scenario to discuss the supersymmetry of the solutions. 
Furthermore, we expect these lower-dimensional models to be useful for other related applications, like the construction of black holes with Lifshitz asymptotics, as well as domain-wall solutions dual to renormalization group flows. Indeed, one of the truncations studied in this paper, based on the $T^{1,1}$ manifold, contains both Lifshitz$_4$ and AdS$_4$ solutions. It would be very interesting to find the interpolating solution between them, describing in the dual theory how conformal invariance is recovered along the flow. 

The paper is organized as follows. The general relation between AdS$_{d+1}$ vacua and Lifshitz$_d$ solutions is proved in section \ref{GeneralResult}.  
In section \ref{SuperLifshitz} we study the supersymmetry properties of the solutions by focusing on the case in which the starting model is a 5-dimensional gauged $\mathcal N=2$ supergravity. In section \ref{ExampleSE} we illustrate these results using a simple supersymmetric consistent truncation of type IIB on arbitrary squashed Sasaki--Einstein manifolds. Another example of consistent truncation, involving the $T^{1,1}$ manifold, is presented in section \ref{T11extension}, and a new non-supersymmetric AdS$_4$ vacuum of type IIB supergravity is found. We conclude in section \ref{Conclusion} by proposing further examples and some directions for future developments.

\medskip

{\bf Note added} : At the same time this paper appeared on the arXiv, we received preprint \cite{HPZ}, which has some overlap with our section 3.

%%%%%%%%%%%%%%%%%%%%%%%%%%%%%%%%%%%%%%%%%%%%%%%%%%%%%%%%%%%%%%
\section{Lifshitz solutions from circle reduction with flux}\label{GeneralResult}

Consider a $(d+1)$-dimensional gravity theory that contains an arbitrary set of gauge fields $A^i$, and scalars $\phi^u$. Fermions can also be present, but we ignore them since they are going to vanish in the solution. The model can come from a consistent truncation of 10- or 11-dimensional supergravity, and it may be or may be not supersymmetric. Under reasonable assumptions, like second order in derivatives and gauge invariance, the general bosonic action takes the form\footnote{We denote by a hat the $(d+1)$-dimensional quantities that risk to be confused with $d$-dimensional ones.}
\bea
\hat S\,=\,\frac{1}{2\kappa_{d+1}^2}\int_{M_{d+1}}\!\!\Big[\,\hat R*1-G_{uv}(\phi)\hat D\phi^u\wedge*\hat D\phi^v-M_{ij}(\phi)\,\hat F^i\wedge*\hat F^j
-\,2\,\hat V(\phi)*1 \,\Big]+\,
\hat S_{\rm top}\,,&& \nn \\ 
\label{d+1DimAction}
\eea
where $\hat V(\phi)$ is the scalar potential, while $\hat F^i=d\hat A^i$ are the field strengths of an Abelian gauge group of arbitrary rank, under which the scalar covariant derivatives are 
\begin{equation}\label{ScalCovDer}
\hat D \phi^u\,=\,d \phi^u + k^u_i(\phi)\,\hat A^i\,.
\end{equation}
Gauge invariance requires the $k_i \equiv k^u_i \,\frac{\partial\,}{\partial \phi^u}$ to be Killing vectors on the scalar manifold.
We have left the topological term unspecified because its particular form depends on the dimension. Indeed, for an Abelian theory in even dimension it is given simply by a combination of field strengths with scalar-dependent coefficients, schematically
\be\label{topeven}
\hat S_{\rm top}\sim\int \,c(\phi)\,\hat F\wedge{\rm\dots}\wedge \hat F\,,
\ee
while in odd dimension it contains a vector potential, but the coefficients are constant
\be\label{topodd}
\hat S_{\rm top}\sim \,c\int \, \hat A\wedge \hat F\wedge{\rm\dots}\wedge \hat F\,.
\ee
Thus, the form of the topological terms is dictated by dimensionality and gauge invariance. In any case, they will not play a relevant role in our results. 

We also assume that the scalar potential $\hat V(\phi)$ has at least one extremum $\phi_{\rm ex}$, 
\begin{equation}\label{extrema}
\frac{\partial \hat V}{\partial\phi^u}\,(\phi_{\rm ex})=0\,,
\end{equation}
with $\hat V(\phi_{\rm ex})<0$. So the theory has at least an AdS$_{d+1}$ solution for constant scalars and vanishing gauge fields.

Finally, we require that the action (\ref{d+1DimAction}) has an additional global symmetry under a transformation of the type $\delta \phi^u = k^u(\phi)$, so in particular $k^u$ has to be a Killing vector generating an isometry of the scalar manifold. Under a redefinition of the scalar fields, we can assume that this transformation acts as a shift symmetry of just one of the scalars $\phi^u$, that we will call $\xi$. Then the Killing vector reads $k\equiv k^u\frac{\partial}{\partial \phi^u}=n \frac{\partial}{\partial \xi}\,$, with $n$ a constant whose role will become clear shortly. In other words, we are requiring that our model contains an axion. In particular, this cannot enter in the scalar potential. We furthermore assume that the axion is uncharged prior to the reduction, i.e. the vectors on the scalar manifold $k^u_i$ appearing in (\ref{ScalCovDer}) have vanishing component along the $\xi$ direction.

The presence of axions is a generic feature of lower-dimensional models derived from string theory, where they typically arise by reducing the higher-dimensional form fields. For instance, suppose one is compactifying a 10- or 11-dimensional supergravity theory containing a $p$-form potential $C_p$ on a manifold having a $p$-cycle $\Sigma_p$, with cohomology representative $\omega_p$. Then, we can expand $C_p= \xi\,\omega_p +$ other terms. Since $\omega_p$ is closed, in the reduced theory the scalar $\xi$ will turn out to be an axion because it does not appear naked in the equations of motion, which are expressed in terms of the field strength $F_{p+1}=dC_p$\,. 

Given these assumptions, in the following we show that there exists a $d$-dimensional Lifshitz solution with $z=2$, precisely at the extremum $\phi_{\rm ex}$ of the $(d+1)$-dimensional scalar potential. To prove this we will reduce the model on a circle $S^1$, also introducing a flux of the axion field strength along it; this will give mass to the vector supporting the solution and in addition will modify the scalar potential. When the $(d+1)$-dimensional model comes from a consistent truncation of 10- or 11-dimensional supergravity, this generalizes the results of \cite{DonosGauntlett_Lif} since the internal manifold is not necessarily Einstein (explicit examples are described in the next sections). The interesting cases for condensed matter applications are $d=4$ and $d=3$.

We take the following ansatz for the $(d+1)$-dimensional metric
\begin{equation}\label{metric}
ds^2(M_{d+1}) =e^{-\frac{2}{d-2}T}\,ds^2(M_d)+e^{2T}\,(d\vartheta+\mathcal A)\otimes(d\vartheta+\mathcal A)\,,
\end{equation}
where $\vartheta$ is the coordinate on $S^1$, while $T$ and $\mathcal A$ are respectively a scalar and a 1-form on the $d$-dimensional spacetime $M_d$. The 1-form $\mathcal A$ transforms as a U(1) gauge field under reparameterizations of the circle coordinate.
Moreover, we expand the $(d+1)$-dimensional gauge fields as
\begin{equation}\label{vector}
\hat A^i\,=\, A^i +\alpha^i\,(d\vartheta+\mathcal A)\,,
\end{equation}
with $A^i$ and $\alpha^i$ being respectively 1-forms and scalars on $M_d$. The expansion in the vielbein $(d\vartheta+\mathcal A)$ rather than just $d\vartheta$ ensures that the $A^i$ are neutral under reparameterizations of $S^1$, so they simply inherit the abelian gauge transformation of the $\hat A^i$. While we assume that the other scalars depend just on the $M_d$ coordinates, for the axion we take 
\be\label{axionansatz}
\xi(x,\vartheta) \,=\, \xi(x) + n\,\vartheta\,,
\ee 
where $x$ collectively denotes the coordinates on $M_d$ and $n$ is a constant.  Hence the $(d+1)$-dimensional derivative of the axion contains a flux term $n\, d\vartheta$ along $S^1$. This induces a new gauging by $\mathcal A$ in the $d$-dimensional covariant derivative. Indeed, under our ansatz the $(d+1)$-dimensional scalar covariant derivatives (\ref{ScalCovDer}) become
\begin{eqnarray}\label{scalar}
\hat D \phi^u &=& d\phi^u+ k^u\,d\vartheta + k^u_j(\phi)\,\left[A^j +\alpha^j\,(d\vartheta+\mathcal A)\right]\nonumber\\[2mm]
&=&D\phi^u+\,\left[k^u + k_j^u(\phi)\,\alpha^j\right](d\vartheta+\mathcal A)\,,
\end{eqnarray}
where $k^u$ is the Killing vector generating the axionic symmetry introduced above, and the $d$-dimensional covariant derivatives read 
\begin{equation}\label{CovDerPhi}
D\phi^u \,=\, d\phi^u- k^u\mathcal A + k^u_j(\phi)\,A^j,
\end{equation}
with, in particular, $D\xi=d\xi-n\,\mathcal A$. So the axion is now St\"uckelberg-coupled to $\mathcal A$, and $n$ is its charge. The last term in (\ref{scalar}) instead contributes to the $d$-dimensional scalar potential. 

When the model comes from a consistent truncation of string theory, there is also a higher-dimensional picture of this new gauging. For an axion originating from a $p$-form potential $C_p$ as sketched above, it corresponds to introduce a flux term $F^{\rm flux}_{p+1} = n\,d\vartheta\wedge\omega_p$ threading $S^1\times \Sigma_p$, in such a way that the field strength $F_{p+1}$ reads
\be
F_{p+1}\,\,=\,\, dC_p+\,F^{\rm flux}_{p+1}  \,\,=\,\, D\xi\wedge \omega_p +\,n\,(d\vartheta+\mathcal A)\wedge \omega_p +\dots\,.
\ee
We will see this mechanism at work in the examples based on consistent truncations of 10-dimensional supergravity presented in the second part of the paper.

The expressions (\ref{metric})--(\ref{CovDerPhi}) can also be seen at the ($d+1$)-dimensional level as an ansatz to solve the $(d+1)$-dimensional equations. We find however more instructive to recast our problem in a $d$-dimensional setup. The reduction on the circle of the action (\ref{d+1DimAction}) can easily be performed and gives the following $d$-dimensional action
\begin{eqnarray}
S\!&=&\!\frac{1}{2\kappa_d^2}\,\int_{M_d}\Big[R*1-\tfrac{d-1}{d-2}\,dT\wedge*dT-e^{-2T}M_{ij}\,d\alpha^i\wedge*d\alpha^j-G_{uv}\,D\phi^u\wedge*D\phi^v\nonumber\\[2mm]
&&\qquad\qquad -\,\frac12\,e^{\frac{2(d-1)}{d-2}T}\,\mathcal F\wedge*\mathcal F-e^{\frac{2}{d-2}T}\,M_{ij}\,(F^i+\alpha^i\,\mathcal F)\wedge*(F^j+\alpha^j\,\mathcal F)\nonumber\\[3mm]
&&\qquad\qquad -  \left(2\,e^{-\frac{2}{d-2}T}\,\hat V + e^{-\frac{2(d-1)}{d-2}T}\,G \right) *1  \Big]\,+\,
S_{\rm top}\,,
\end{eqnarray} 
where $F^i=dA^i$ and $\mathcal F=d\mathcal A$, while $\kappa_d^{-2}=\kappa_{d+1}^{-2}  \int_{S^1} d\vartheta$. Notice in particular the new contribution to the scalar potential
\be
G(\phi,\alpha)\,:=\, G_{uv}\left[ k^u + k_i^u(\phi)\,\alpha^i\right]\left[k^v + k_j^v(\phi)\,\alpha^j\right].
\ee

If the $(d+1)$-dimensional action is a supergravity model, the $d$-dimensional action will also be supersymmetric, with the same amount of supercharges and the additional gauging of the axionic symmetry. Moreover, if the former arises as a consistent truncation of higher-dimensional supergravity, the latter will also correspond to a consistent truncation, since the circle reduction preserving the U(1) singlets does not spoil consistency. Note from (\ref{axionansatz}) that actually the axion is not such a singlet, however the fact that it always appears in the equations through its derivative, which instead is U(1) invariant, ensures consistency.

We now show our main result: {\it the $d$-dimensional theory has a $z=2$ Lifshitz solution of the form (\ref{LifMetric}) at each extremum of the $(d+1)$-dimensional scalar potential with negative value, that is, for each AdS$_{d+1}$ solution of the $(d+1)$-dimensional theory.}

The vielbeine of the Lifshitz metric (\ref{LifMetric}) are
\begin{equation}\label{VielbLifshitz}
\theta^t = L\, r^z\,dt\;,\hspace{1.5cm}\theta^r= L\,\frac{dr}{r} \;,\hspace{1.5cm}
\theta^\ell =L\, r\,dx^\ell\,,
\end{equation}
and our orientation choice is $\,*1 = \theta^t\wedge \theta^1\wedge \ldots\wedge \theta^{d-2}\wedge \theta^r$.
In flat indices, the non-vanishing components of the Ricci tensor are
\begin{equation}
R_{tt}=  \frac{z\,(z+ d-2 )}{L^2}\,,\qquad\quad
R_{\ell\ell}=-\frac{z+ d-2}{L^2}\,,\qquad\quad
R_{rr}= -\frac{z^2+ d-2}{L^2}\,.\label{RicciTensor_Lif}
\end{equation}
We take all the $d$-dimensional scalars to be constant, and the gauge field coming from the reduction of the metric of the form
\begin{equation}\label{AnsatzMathcalA}
\mathcal A=\lambda\,\theta^t,
\end{equation}
with $\lambda$ a constant, which without loss of generality we assume positive. Moreover, we fix the other gauge fields to
\begin{equation}\label{RelAmathcalA}
A^i=-\alpha^i\,\mathcal A\,.
\end{equation}
As a consequence, we have $F^i+\alpha^i\,\mathcal F=0$ and $D\phi^u=- \left(k^u + k_j^u \alpha^j\right) \mathcal A$. 

Now, the equations of motion for $A^i$ and $\alpha^i$ are all satisfied if we impose\footnote{Of course the contributions of the topological terms have to vanish. This is the case as one can easily see by reducing (\ref{topeven}) and (\ref{topodd}): all the pieces in the equations of motion turn out to be proportional to $d\alpha$ and/or $F+\alpha\,\mathcal F$ and both are zero within our ansatz.}
\begin{equation}\label{VectorEquations}
k_i^uG_{uv} k_j^v \,\,\alpha^j  \,=\, -\,k_i^u G_{uv}k^v  \,.
\end{equation}
This is a set of linear equations for the $\alpha^i$, which can always be solved. If the matrix $k_i^uG_{uv}k_j^v$ has maximal rank then all the $\alpha^i$ are fixed, otherwise some of them will remain moduli of the solution. In the next sections we will see examples of both situations.

Next, within our ansatz the equations for the scalars $\phi^u$ boil down to 
\begin{equation}
e^{-\frac{2}{d-2}T}\,\frac{\partial \hat V}{\partial\phi^u} \,+\,  \frac12 \left(e^{-\frac{2(d-1)}{d-2}T}- \lambda^2\right) \frac{\partial G}{\partial\phi^u} \,= \,0\,.
\end{equation}
By fixing
\begin{equation}\label{RelTlambda}
e^{-\frac{d-1}{d-2}T}\,=\, \lambda\,, 
\end{equation}
these are then solved at any extremum of the $(d+1)$-dimensional scalar potential $\hat V$.
It is clear that if any of the scalars $\phi^u$ does not appear in the potential, its value will remain undetermined in the Lifshitz solution as well. 

The equation for the gauge field $\mathcal A$ reads
\begin{equation}\label{eq_for_L}
\frac{(d-2)\,z}{L^2}\, =\, 2\, e^{-\frac{2(d-1)}{d-2}T}\,G(\phi_{\rm ex}, \alpha)\,,
\end{equation}
where $G$ is positive due to the positive-definiteness of the scalar metric $G_{uv}$. Continuing, the equation of $T$ yields
\begin{equation} \label{eq_for_lambda}
\frac{z^2+(d-2)\,z}{L^2} \,+\, \frac{4}{d-1}\,e^{-\frac{2}{d-2}T}\,\hat V(\phi_{\rm ex}) \,\,=\,\, 0\,.
\end{equation}
Finally, we are left with the Einstein equation, which in flat spacetime indices $a,b$ and using the relations coming from the matter equations of motion, reads
\bea
R_{ab}\!\!&=&\!\! G\, \mathcal A_a \mathcal A_b  + \frac12\,e^{\frac{2(d-1)}{d-2}T}\iota_a\mathcal F\lrcorner\,\iota_b\mathcal F - \frac{\eta_{ab}}{2(d-2)}\,\Big[e^{\frac{2(d-1)}{d-2}T}\mathcal F\lrcorner\,\mathcal F - 4\,e^{-\frac{2}{d-2}T}\hat V \,-\, 2\,e^{-\frac{2(d-1)}{d-2}T}\,G\Big] \nn\\ [2mm]
\!\! &=&\!\!\frac{1}{2L^2}\Big[\big(z^2 +(d-2)\,z\big)\,\delta_a^t \delta_b^t -z^2\,\delta_a^r \delta_b^r - \,\eta_{ab}\big(z^2+(d-2)\,z\big)\Big] .
\eea
Recalling (\ref{RicciTensor_Lif}), one can see that all components are satisfied provided we take $z=2$. Plugging this back in (\ref{eq_for_L}) and (\ref{eq_for_lambda}), we obtain the values of the coefficient $\lambda$ in the vector ansatz and the size $L$ of the Lifshitz spacetime:
\be\label{EqsLlambda}
L^2\lambda^2 \,=\, \frac{d-2}{G}\,,\qquad\quad \qquad L^2\lambda^{\frac{2}{d-1}} \,=\, -\frac{d\,(d-1)}{2\hat V}\,.
\ee
Note that this makes sense only if $\hat V(\phi_{\rm ex})<0$, namely if the extremum of $\hat V$ corresponds to an AdS$_{d+1}$ solution. The quantity $-\frac{d\,(d-1)}{2\hat V}$ is precisely the square radius of AdS$_{d+1}$.

%%%%%%%%%%%%%%%%%%%%%%%%%%%%%%%%%%%%%%%%%%%%%%%%%%%%%%%%%%%%%%%
\section{Supersymmetric Lifshitz backgrounds}\label{SuperLifshitz}

In this section we fix $d=4$ and study the supersymmetry conditions for a Lifshitz background, working in the context of $\mathcal N=2$ supergravity. We show that if one starts from an AdS$_5$ solution which preserves $\mathcal N=2$ supersymmetry, then the associated Lifshitz$_4$ solution constructed as above preserves 1/4 of the supercharges, and we find the explicit Killing spinors. A summary of our results is given at the end of subsection \ref{AdStoLifSusyBreaking}.

\subsection{Preliminary study of the $d=4$, $\mathcal N=2$ susy variations}

We start by writing down a set of general conditions for supersymmetric Lifshitz backgrounds in 4-dimensional $\mathcal N=2$ supergravity.

The bosonic sector of the gauged $\mathcal{N}=2$ supergravity action coupled to vector and hypermultiplets has general structure\footnote{We refer to \cite{N=2review} for a comprehensive review of 4-dimensional $\mathcal N=2$ supergravity. Our conventions are given in appendix \ref{conventions}.}
\begin{equation}\label{4dN=2action}
   S = \!\int \tfrac{1}{2}R* 1
            \,+ \tfrac{1}{2}{\rm Im}\mathcal{N}_{IJ}F^I \wedge * F^J
      -\, \tfrac{1}{2}{\rm Re}\mathcal{N}_{IJ}F^I\wedge F^J
      -\, g_{i\bar\jmath}Dz^i\wedge* D\bar{z}^{\bar \jmath}
      \,-\, h_{uv}Dq^u\wedge* Dq^v 
      - V *1.
\end{equation}
In addition to the metric and the graviphoton $A^0$ in the gravity multiplet, this contains $n_H$ hypermultiplets with $q^u$ real scalars, $u=1,\ldots,4n_H$, parameterizing a quaternionic-K\"ahler manifold with metric $h_{uv}$. There are also $n_V$ vector multiplets $(A^i, z^i)$, $i=1,\ldots,n_V$, where the $A^i$ are 1-forms and the $z^i$ complex scalars. Including the graviphoton, the 1-forms are collectively denoted by $A^I,$ $I=0,1,\ldots, n_V$, and $F^I$ are their field strengths. The vector multiplet scalar manifold is a special K\"ahler space with metric $g_{i\bar\jmath}\,$, derived from the K\"ahler potential $K$ as $g_{i\bar\jmath} = \partial_i\partial_{\bar\jmath}K$. In turn, the K\"ahler potential is determined by the holomorphic symplectic section $(X^I, \mathcal F_I)$ on the special K\"ahler manifold via
\be\label{FormulaK}
K\,=\, -\log \left[i\,(\bar X^I \mathcal F_I -  X^I\bar{ \mathcal F_I}) \right] .
\ee
When a prepotential function $\mathcal F(X)$ exists,\footnote{This $\mathcal F(X)$ should not be confused with the field strength $\mathcal F=d\mathcal A$ appearing in the previous section.} one has $\mathcal F_I = \frac{\partial \mathcal F}{\partial X^I}$. The period matrix $\mathcal N_{IJ}$ governing the gauge kinetic terms follows from the formula
\begin{equation}\label{FormulaForN}
\mathcal N_{IJ} \,=\, \overline{\mathcal F}_{IJ} \,+\, 2i\, \frac{({\rm Im}\, \mathcal F_{IK})X^K  ({\rm Im}\, \mathcal F_{JL})X^L}{X^M ({\rm Im}\,\mathcal F_{MN})X^N}\,,\qquad\qquad \mathcal F_{IJ} := \frac{\partial^2\mathcal F}{\partial X^I\partial X^J}\,.
\end{equation}

We will consider an electric gauging of the quaternionic isometries, with no gauging of the special K\"ahler isometries. Then the gauge group is Abelian (so $F^I=dA^I$), and the scalar covariant derivatives read $Dz^i \equiv dz^i$ and 
\be\label{generalDq}
Dq^u = dq^u + k^u_I A^I,
\ee 
where $k_I = k^u_I\frac{\partial}{\partial q^u}$ are the Killing vectors generating the quaternionic isometries being gauged.
With this gauging, the general expression for the $\mathcal N=2$ scalar potential is
\begin{equation}\label{4dN=2ScalPot}
V\,=\, 4 \,e^{K} h_{uv} k^u_I k^v_J X^I \bar X^J - \left[\tfrac{1}{2}({\rm Im}\,\mathcal N)^{-1\,IJ} + 4\, e^K X^I\bar X^J\right] P^x_I P^x_J\,,
\end{equation}
where the $P^x_I$, $x=1,2,3$, are the triplet of Killing prepotentials associated with the gauged isometries. They are defined by
\begin{equation}\label{DefPx}
\iota_{k_I}\Omega^x = dP^x_I + \epsilon^{xyz}\omega^yP^z_I\,,
\end{equation} 
where $\omega^x$ is the connection of the SU(2) bundle existing on any quaternionic manifold, and
\be
\Omega^x \,:=\, d\omega^x +\frac{1}{2}\epsilon^{xyz}\omega^y\wedge \omega^z
\ee
is the associated curvature.

We are interested in a purely bosonic background with constant values of the scalars and non-vanishing vectors. To have a supersymmetric configuration, we thus need to impose the vanishing of the variations of the fermionic fields. These are the gravitini $\psi_{A\mu}$, the gaugini $\lambda^{iA}$ and the hyperini $\zeta_\alpha$, where $A=1,2$ is the SU(2) R-symmetry index and $\alpha =1,\ldots, 2n_H$ is an Sp$(n_H)$ index. The general supersymmetry variations of the positive-chirality fermionic fields given in \cite{N=2review} reduce to
\begin{eqnarray}
\delta\psi_{A\mu} &=& \nabla_\mu\vep_A \,+\, \tfrac i2 A_\mu^I P^x_I (\sigma^x)_A{}^B \vep_B  \,+\,  e^{\frac{K}{2}}\im \,\mathcal N_{IJ}X^I F^{J-}_{\mu\nu}\gamma^\nu\epsilon_{AB} \vep^B  \,-\, S_{AB} \gamma_\mu\vep^B , \nonumber \\ [2mm]
\delta\lambda^{iA} &=& \tfrac 12 e^{\frac{K}{2}} g^{i\bar\jmath} \bar D_{\bar\jmath}\bar X^{K} \im \,\mathcal N_{KL} F_{\mu\nu}^{L-}\gamma^{\mu\nu}\epsilon^{AB}\vep_B \,+\, W^{iAB}\vep_B\,,\nonumber\\ [2mm]
\delta\zeta_\alpha &=&  \,\mathcal U_u^{B\beta} k_I^u A_\mu^I  \gamma^\mu \vep^A\epsilon_{AB}\mathbb C_{\alpha\beta} \,+\, N^A_\alpha \vep_A\,. \label{eq:N=2susyVariations}
\end{eqnarray}
Here, $\nabla_\mu$ is the usual Lorentz covariant derivative on spinors, $\nabla_\mu\vep = \partial_\mu\vep +\frac{1}{4}\omega_\mu^{ab}\gamma_{ab}\vep$, while 
$F_{\mu\nu}^{I-} \,:=\, \tfrac{1}{2}\left( F^I_{\mu\nu} - \tfrac{i}{2}\epsilon_{\mu\nu\rho\sigma} F^{I\rho\sigma} \right)$ are anti self-dual field strengths, and $D_j X^I := (\partial_j + \partial_j K) X^I$ is the K\"ahler covariant derivative of the holomorphic section $X^I(z)$ on the special K\"ahler manifold.
The fermionic shifts determined by the gauging are given by
\begin{eqnarray}
S_{AB} &=& \tfrac{i}{2}\,e^{\frac{K}{2}}(\sigma_x)_{AB} P^x_I X^I ,\nn \\ [2mm]
W^{iAB} &=& i \,e^{\frac{K}{2}} (\sigma_x)^{AB} P^x_L \,g^{i\bar\jmath}\bar D_{\bar\jmath}\bar X^{L} ,\nn \\ [2mm]
N^A_\alpha &=& 2\,e^{\frac{K}{2}}\,\mathcal U^{\;A}_{u\,\alpha} \, k^u_I \bar X^I \,.\label{FermionicShifts}
\end{eqnarray}
The quaternionic vielbein $\mathcal U^{A\alpha}=\mathcal U_u^{A\alpha}(q)dq^u$ satisfies 
\begin{equation}
h_{uv}dq^udq^v = \mathbb C_{\alpha \beta} \epsilon_{AB}\mathcal U^{A\alpha} \mathcal U^{B\beta}
\end{equation}
as well as the reality condition
\begin{equation}\label{realityU}
\left( \mathcal U^{A\alpha}\right)^* \,=\, \epsilon_{AB} \mathbb C_{\alpha\beta} \mathcal U^{B\beta}\,,
\end{equation}
$\epsilon_{AB}$ and $\mathbb C_{\alpha\beta}$ being the antisymmetric metrics of SU(2)\,$\cong$\,Sp(1) and Sp$(n_H)$, respectively.
More details on our conventions are given in appendix \ref{conventions}.

In the following we work in flat spacetime indices, using the vielbeine (\ref{VielbLifshitz}) of the Lifshitz metric. We fix an ansatz in which all vectors are parallel,
\be\label{ParallelVectors}
A^I \,=\, A^I_t \,\theta^t \qquad \Rightarrow \qquad F^I = dA^I = A^I_t \,\frac{z}{L} \, \theta^r\wedge \theta^t \,,
\ee
where the $A^I_t$ are assumed to be constant. Then the non-vanishing components of $F^{I-}$ are
\be
F^{I-}_{tr} = -\frac{z}{2L}A^I_t\;,\qquad\quad F^{I-}_{12} = -\frac{iz}{2L}A^I_t\qquad\Rightarrow\qquad  F^{I-}_{\mu\nu}\gamma^{\mu\nu} = \frac{z}{L} A^I_t \gamma^r\gamma^t (\mathbbone+\gamma_5)\,.
\ee
Plugging the expressions (\ref{FermionicShifts}) for the fermion shifts in, the hyperino equation $\delta\zeta_\alpha=0$ gives
\begin{equation}\label{HypEq}
\mathcal U_u^{A\alpha}k^u_I \left( A_t^I  \gamma^t \vep^B\epsilon_{BA} \,+\, 2\,e^{\frac{K}{2}} \bar X^I \vep_A\right) \,=\,0\,,
\end{equation}
while the gaugino equation $\delta \lambda^{iA}=0$ is
\be\label{GauginoEqBis}
\bar D_{\bar\jmath}\bar X^{I} \left(\,\frac{z}{L}\, \im \,\mathcal N_{IK} A^K_t \gamma^r\gamma^t \vep_A \,-\, i \, P^x_I(\sigma_x)_A{}^{B}  \vep_B \,\right)\,=\,0\,.
\ee
Evaluating the spin connection of the Lifshitz metric, the gravitino equation $\delta\psi_{A\mu} = 0$ yields
\begin{eqnarray}
\partial_t \vep_A - \tfrac{z}{2L}\gamma^t\gamma^r \vep_A + \tfrac{i}{2} A^I_t P^x_I (\sigma^x)_A{}^B\vep_B - \tfrac{z}{2L} e^{\frac{K}{2}} {\rm Im}\mathcal N_{IJ} X^I A^J_t \gamma^r \epsilon_{AB} \vep^B + S_{AB} \gamma^t \vep^B \!\!&=&\!\! 0,\nonumber\\ [2mm]
\partial_\ell \vep_A +\tfrac{1}{2L}\gamma^\ell\gamma^r \vep_A  + \tfrac{z}{2L} e^{\frac{K}{2}} {\rm Im}\mathcal N_{IJ} X^I A^J_t\, \gamma^\ell \gamma^t \gamma^r \epsilon_{AB} \vep^B  - S_{AB} \gamma^\ell \vep^B \!\!&=&\!\! 0 ,\nonumber \\[2mm]
\partial_r \vep_A  + \tfrac{z}{2L} e^{\frac{K}{2}} {\rm Im}\mathcal N_{IJ} X^I A^J_t\, \gamma^t \epsilon_{AB} \vep^B - S_{AB} \gamma^r \vep^B \!\!&=&\!\! 0.\qquad\quad\label{GravitEqBis}
\end{eqnarray}

\subsection{1/4 BPS Lifshitz backgrounds}\label{AdStoLifSusyBreaking}

In the spirit of section \ref{GeneralResult}, we now restrict to the situation in which the 4-dimensional theory comes from circle reduction of a 5-dimensional, $\mathcal N=2$ supergravity theory \cite{GunaydinSierraTownsend}. In this case, the 4-dimensional theory has some specific features, which we illustrate in the following.

If the 5d theory has $n_H$ hypermultiplets and $n_V-1$ vector multiplets, then the dimensional reduction yields a 4d theory with the same hypermultiplets and $n_V$ vector multiplets.  
Each vector multiplet of 5-dimensional, $\mathcal N=2$ supergravity is made by a vector and a real scalar. The reduction of the vector provides the missing scalar completing the 4d vector multiplet.
The additional 4d vector multiplet is obtained from the 5d graviphoton, together with the scalar coming from the $\vartheta\vartheta$-component of the metric and controlling the size of the reduction circle; this is $T$ in our metric ansatz (\ref{metric}). The 4d graviphoton $A^0$ comes from the components of the 5d metric with mixed indices $g_{\mu \vartheta} = e^{2T}\mathcal A_\mu$ (one can see that the precise relation is $A^0 = -\frac{1}{\sqrt 2} \mathcal A$).

As in the previous section, we call $\alpha^i$, $\,i=1,\ldots,n_V$, the scalars coming from the reduction of the 5d gauge fields (including the 5d graviphoton). In 5-dimensional, $\mathcal N=2$ supergravity the scalar manifold of vector multiplets is the hypersurface parameterized by real coordinates $h^i$, satisfying
\be
\frac 16\, d_{ijk}h^ih^jh^k \,=\, 1\,,
\ee
where $d_{ijk}$ is a real, symmetric, constant tensor which also specifies the 5d Chern--Simons term. This is called a real special geometry, and the relation determining the corresponding special K\"ahler manifold of 4d vector multiplets is often called the r-map. In the dimensional reduction, the $h^i$ combine with the scalar $T$ from the 5d metric, and give rise to $n_V$ unconstrained scalars $\rho^i: = e^T h^i$. The function 
\be\label{Kfrom5d}
K \,=\, -\log \Big(\frac{4}{3}d_{ijk}\rho^i\rho^j\rho^k \Big) \,=\, -\log \left( 8\,e^{3T} \right),
\ee
is identified as the K\"ahler potential of the special K\"ahler manifold. Defining the complex scalars $z^i=\alpha^i + i \rho^i$, and with the choice of special coordinates $X^I=(X^0,X^i)=(1,z^i)$, this K\"ahler potential can be derived via (\ref{FormulaK}) from a cubic prepotential, of the form 
\be\label{CubicPrepot}
\mathcal F(X) \,=\, -\frac{1}{6}d_{ijk}\frac{X^iX^jX^k}{X^0}\,.
\ee 

Clearly, AdS$_5$ solutions can be found only if the 5-dimensional supergravity theory contains a non-trivial scalar potential, namely if it is gauged (we refer to \cite{CeresoleDallAgataN=2} for gauged $\mathcal N=2$ supergravity in 5 dimensions). We assume a gauging of the quaternionic isometries only, whose Killing vectors $k^u_i$ and respective Killing prepotentials $P^x_i$ are identified with the ones in the 4-dimensional theory introduced above. Along the lines of the previous section, we require that the hyperscalar manifold has an axionic direction parameterized by $\xi$, which we gauge in the circle reduction process by introducing a component along $S^1$, $n \,d\vartheta$, of the axion derivative. This additional gauging involves just the 4-dimensional graviphoton $A^0=-\frac{1}{\sqrt 2} \mathcal A$, and is generated by a Killing vector $k_0 =\sqrt 2\, n\, \frac{\partial}{\partial\xi}\,$, with associated Killing prepotential $P^x_0$. 
So the 4d Killing prepotentials are $P^x_I = (P^x_0,P^x_i)$, with the $P^x_i$ coinciding with their 5d counterparts.

The conditions for an $\mathcal N=2$ AdS$_5$ solution within the 5d theory can be expressed in terms of simple algebraic equations. These are \cite{CDKV_DomWalls}
\be
k^u_i h^i = 0\,,\qquad\qquad P^x_i = \frac 16\, d_{ijk} h^jh^k (P^x_lh^l)
\ee
and can be rephrased in 4d language as
\be\label{SusyAdSin4dLanguage}
k^u_i \rho^i = 0\,,\qquad\qquad P^x_i  + \frac{2i}{3}\,\partial_i K \,(P^x_j\rho^j)\,=\, 0\,,
\ee
where we used the fact that for $K$ of the form (\ref{Kfrom5d}) one has $\partial_i K\equiv \frac{\partial}{\partial z^i} K = 2i \,e^K d_{ijk} \rho^j\rho^k$.

We now analyze the supersymmetry conditions for a Lifshitz background of the 4-dimensional theory described above, assuming that the conditions (\ref{SusyAdSin4dLanguage}) for a supersymmetric AdS$_5$ vacuum are verified.

We specify the vector ansatz (\ref{ParallelVectors}) to
\be\label{AIproptoAlphaI}
A^I_t \,=\, \alpha^I A^0_t\,,
\ee 
where $\alpha^I := \re\, X^I=(1,\alpha^i)$ and $A^0_t$ is a non-vanishing constant. With the identification $A^0_t= -\tfrac{1}{\sqrt 2}\mathcal A_t  = -\frac{1}{\sqrt 2} \lambda$, this corresponds to the ansatz taken in (\ref{RelAmathcalA}) to solve the equations of motion (note that since we assumed $\lambda >0$, here we have $A^0_t<0$).

Using the first condition in (\ref{SusyAdSin4dLanguage}), the hyperino equation (\ref{HypEq}) becomes
\be\label{HypEqElaborated}
\mathcal U_u^{A\alpha}k^u_I\alpha^I \left(  A^0_t\,\gamma^t \vep^B\epsilon_{BA} \,+\, 2\,e^{\frac{K}{2}} \vep_A\right) \,=\,0\,.
\ee
Acting with $A^0_t\gamma^t$, taking the complex conjugation (recalling (\ref{realityU})), and using (\ref{HypEqElaborated}) another time, we arrive at
\be\label{SquareHypEq}
\mathcal U_u^{A\alpha}k^u_I\alpha^I \vep_A \left[  (A^0_t)^2 \,-\, 4\,e^{K} \right] \,=\,0\,.
\ee
Note that the quantity $k^u_I\alpha^I$ cannot vanish, because its component in the direction of the axion $\xi$ is just equal to $n\neq 0$. So to solve the equation we take
\be\label{RelKA0}
2 \, e^{\frac{K}{2}} \,=\, -\, A^0_t\,.
\ee
Recalling the form of $K$ in (\ref{Kfrom5d}), this gives the same relation between the size $T$ of the reduction circle and the temporal component of the massive vector that was found in (\ref{RelTlambda}) while studying the equations of motion.
Then (\ref{HypEqElaborated}) reduces to the projector
\be\label{HypProjector}
\gamma^t \vep^A \,=\, \,\epsilon^{AB}\vep_B\qquad \Leftrightarrow \qquad \vep^2 \,=\, \gamma^t\,\vep_1\,,
\ee
which fixes the second supersymmetry parameter in terms of the first.

Next we turn to the gaugino equation (\ref{GauginoEqBis}). Starting from the cubic prepotential (\ref{CubicPrepot}) and using the general formula (\ref{FormulaForN}), one derives the useful relation
\be\label{Simplification}
{\rm Im}\mathcal N_{IJ} {\rm Re}X^J =\, -\frac{1}{8}\, e^{-K}\,\delta_I^0 \,.
\ee 
Recalling (\ref{AIproptoAlphaI}), this can be used together with (\ref{RelKA0}) to simplify the first term of (\ref{GauginoEqBis}). Using the AdS$_5$ susy condition (\ref{SusyAdSin4dLanguage}) in the second term, we arrive at
\be\label{GauginoEqTer}
\frac{z}{2\,L\,A^0_t}\, \gamma^r\gamma^t \vep_A \,+\,\Big( \frac{1}{3}\,\rho^i P^x_i \pm i\,\alpha^I P^x_I \Big) (\sigma_x)_A{}^{B}  \vep_B \,=\,0\,.
\ee  
The equation with the minus sign is obtained starting from the one with the plus, taking the complex conjugate (which flips the upper/lower position of the $A,B$ indices), multiplying by $\gamma^t$ and eventually using the projector (\ref{HypProjector}).
Subtracting these two equations we have
\be
\alpha^I P^x_I (\sigma_x)_A{}^{B}  \vep_B \,=\,0\,,
\ee
which, for non-vanishing $\vep_A$, requires $\,\det \left(\alpha^I P^x_I \sigma_x\right)=0$, and therefore the condition
\be\label{EqforAlphaSusy}
\alpha^I P^x_I \,\equiv\, P^x_0 + \alpha^i P^x_i \,=\, 0\,.
\ee
Plugging this back in (\ref{GauginoEqTer}) gives
\be\label{GauginoEq4}
\frac{z}{2\,L\,A^0_t}\, \gamma^r\gamma^t \vep_A \,+\, \frac{1}{3}\,\rho^i P^x_i (\sigma_x)_A{}^{B}  \vep_B \,=\,0\,,
\ee 
which upon multiplication by $\frac{z}{2\,L\,A^0_t} \,\gamma^r\gamma^t$ yields, for non-vanishing $\vep_A$,
\be\label{IntegrabilityGauginoEq}
\Big(\frac{z}{2\,L\,A^0_t}\Big)^2  \,=\, \frac{1}{9} (\rho^i P^x_i) (\rho^j P^x_j) \,.
\ee
When this is satisfied, eq. (\ref{GauginoEq4}) is just a projector. Though we have two equations, labeled by $A=1,2$, using the hyperino projector (\ref{HypProjector}) it is easy to see that one implies the other. 
Hence the hyperino and the gaugino equations together give rise to two projections. 

Finally, we study the gravitino equation (\ref{GravitEqBis}). Using (\ref{Simplification}) and recalling (\ref{RelKA0}) we have 
\be
e^{\frac{K}{2}} {\rm Im}\mathcal N_{IJ} X^I A^J_t \,=\, -\frac{1}{8}e^{-\frac{K}{2}}A^0_t \,=\, \frac{1}{4} \,.
\ee 
Also noting that (\ref{AIproptoAlphaI}), (\ref{EqforAlphaSusy}) set $A^I_t P^x_I=0$, the gravitino equation in flat indices becomes
\begin{eqnarray}
\partial_t \vep_A - \tfrac{z}{2L}\gamma^t\gamma^r \vep_A - \tfrac{z}{8L} \,\gamma^r \epsilon_{AB} \vep^B + S_{AB} \gamma^t \vep^B \!\!&=&\!\! 0\,,\nonumber\\ [2mm]
\partial_\ell \vep_A +\tfrac{1}{2L}\gamma^\ell\gamma^r \vep_A  + \tfrac{z}{8L} \, \gamma^\ell \gamma^t \gamma^r \epsilon_{AB} \vep^B  - S_{AB} \gamma^\ell \vep^B \!\!&=&\!\! 0 \,,\nonumber \\[2mm]
\partial_r \vep_A  + \tfrac{z}{8L}\, \gamma^t \epsilon_{AB} \vep^B - S_{AB} \gamma^r \vep^B \!\!&=&\!\! 0\,.
\end{eqnarray}
Noting that (\ref{EqforAlphaSusy}) reduces the gravitino shift in (\ref{FermionicShifts}) to
$S_{AB} = -\tfrac{1}{2}\, e^{\frac{K}{2}}(\rho^iP^x_i)(\sigma_x)_{AB} \,,
$ and using the projectors (\ref{HypProjector}), (\ref{GauginoEq4}) from the hyperino and gaugino equations, \hbox{we arrive at}
\begin{eqnarray}
\partial_t \vep_A  \!\!&=&\!\! 0\,,\nonumber\\ [2mm]
\partial_\ell \vep_A +\tfrac{2-z}{4L}\,\gamma^\ell\gamma^r \vep_A  \!\!&=&\!\! 0 \,,\nonumber \\[2mm]
\partial_r \vep_A  \,-\, \tfrac{z}{2L} \vep_A \!\!&=&\!\! 0\,.
\end{eqnarray}
The second equation ($\ell=1,2$) gives the integrability condition
\be
0\,=\, \tfrac 12(\partial_1\partial_2 - \partial_2\partial_1)\vep_A \,=\, \left(\tfrac{2-z}{4L}\right)^2 \gamma^1\gamma^2\vep_A \,,
\ee
which imposes $z=2$. Hence, after reinstating curved indices the Killing spinor solution is
\be\label{KillingSpinor}
\vep_A \,=\,\, r\:\widetilde\vep_A\,,
\ee
where $\widetilde \vep_A$ is a constant spinor satisfying the projections (\ref{HypProjector}), (\ref{GauginoEq4}).

\subsubsection*{Summary and comments}

We have solved the conditions for a supersymmetric Lifshitz background in the context of 4-dimensional gauged $\mathcal N=2$ supergravity coupled to an arbitrary number of vector multiplets and hypermultiplets. This is assumed to descend from a 5-dimensional theory admitting an $\mathcal N=2$, AdS$_5$ vacuum.

The Killing spinor solution is $\vep_A \,=\,\, r\:\widetilde\vep_A$, where the constant spinors $\widetilde \vep_A$ satisfy the two projections (\ref{HypProjector}), (\ref{GauginoEq4}) from the hyperino and the gaugino equations.
Then the Lifshitz solution is 1/4 BPS, namely two supercharges are preserved.

The dynamical exponent of the Lifshitz metric is $z=2$. In addition to conditions (\ref{SusyAdSin4dLanguage}) for an $\mathcal N=2$ AdS vacuum in 5 dimensions, the bosonic fields satisfy relations (\ref{AIproptoAlphaI}), (\ref{RelKA0}), (\ref{EqforAlphaSusy}) and (\ref{IntegrabilityGauginoEq}). These are compatible with the outcome of the study of the equations of motion in the previous section. Indeed, as it is well known the conditions for an $\mathcal N=2$ AdS vacuum guarantee extremization of the 5d potential. Moreover, as already remarked, relations (\ref{AIproptoAlphaI}), (\ref{RelKA0}) correspond to (\ref{RelAmathcalA}), (\ref{RelTlambda}).
We also note that for $\mathcal N=2$ AdS$_5$ backgrounds, the term on the right hand side of (\ref{IntegrabilityGauginoEq}) is proportional to the on-shell value of the 5d scalar potential $\hat V$ (in our conventions, the relation is $\hat V = -\frac{1}{3}(h^iP^x_i)(h^jP^x_j)\,$). Hence this equation corresponds to the second in (\ref{EqsLlambda}), fixing the relation between the 5d potential and the Lifshitz radius $L$.
However, eqs. (\ref{VectorEquations}) and (\ref{EqsLlambda}), coming from the vector equations of motion, seem not to follow from the supersymmetry conditions and should therefore be imposed separately. On the other hand, supersymmetry requires the extra condition (\ref{EqforAlphaSusy}). Depending on the specific model under study, this might or might not further constrain the scalars $\alpha^i$. However, (\ref{EqforAlphaSusy}) is not necessarily in contradiction with (\ref{VectorEquations}), because in the supersymmetric case, due to condition $k^u_i\rho^i=0$, the matrix $k_i^uG_{uv}k_j^v$ is not invertible, meaning that not all the $\alpha$'s are fixed by the equations of motion.

Explicit examples of supersymmetric Lifshitz solutions are given in the next sections.

%%%%%%%%%%%%%%%%%%%%%%%%%%%%%%%%%%%%%%%%%%%%%%%%%%%%%%%%%%%%%%
\section{$\!\!\!$A simple example from squashed Sasaki-Einstein spaces}
\label{ExampleSE}

In this section, we illustrate our general results fixing $d=4$, which is the most interesting dimension for condensed matter applications as originally proposed in \cite{Kachru}. We do this by providing a consistent truncation of type IIB supergravity on the direct product of $S^1$ with any 5-dimensional squashed Sasaki--Einstein manifold. As we explicitly show, the model is compatible with gauged $\mathcal N=2$ supergravity.  
Prior to the $S^1$ reduction, the model contains two AdS$_5$ vacua: one is $\mathcal N=2$ supersymmetric, so that a 1/4 BPS Lifshitz$_4$ solution is obtained, while the other provides a new, non-supersymmetric embedding of Lifshitz geometry into string theory.

\subsection{The consistent truncation}

We consider the 5-dimensional gauged $\mathcal N=2$ supergravity coupled to the universal hypermultiplet obtained as a consistent truncation of type IIB supergravity on 5-dimensional squashed Sasaki--Einstein manifolds \cite{IIBonSE}. This represents one of the simplest supersymmetric models containing an axion. The bosonic action reads\footnote{For any $p$-form $\varphi$ we use the shorthand notation $\varphi^2\equiv\varphi\lrcorner\varphi$, with the index contraction $\lrcorner$ including the $\frac{1}{p!}$ factor.}
\begin{eqnarray}\label{5duniversal}
S \!&=&\!\frac{1}{2\kappa_5^2}\,\int\,\bigg[\,\Big(\,\hat R-\frac{1}{2}\,\cosh^2{\sigma}\,d\phi^2-\frac{1}{2}\,e^{2\phi}\cosh^4{\sigma}\,(dC_0)^2-2\,d\sigma^2- \,\frac{1}{2}\,\sinh^2{(2\sigma)}\,\hat D\chi^2\quad\nonumber\\[2mm]
&&\quad\quad\quad\quad +\,\frac{1}{2}\,e^\phi\,\sinh^2{(2\sigma)}\,dC_0\lrcorner\hat D\chi- \frac{3}{2}\,d\hat A^2-\, 2\, \hat V\Big) *1+\hat A\wedge d\hat A\wedge d\hat A \,\bigg]\,,
\end{eqnarray}
where the potential is 
\begin{equation}
\hat V\,=\, \frac{3}{2}\,\cosh^2{\sigma}\big[\cosh{(2\sigma)}-5\big] .
\end{equation}
Here, $C_0$ is the type IIB axion and $\phi$ is the dilaton. The vector $\hat A$ is the one gauging the Reeb isometry of the Sasaki--Einstein manifold and we have a scalar, $\chi$, charged under it: $\hat D\chi=d\chi-3\,\hat A$. This can be seen as the phase of a complex scalar, whose modulus is a function of $\sigma$.\footnote{Thanks to the presence of a charged scalar, (a non-supersymmetric subsector of) this model already proved to be interesting for condensed matter applications: it provides the embedding into string theory of a holographic superconductor \cite{Gubser_sup}.} Notably, the potential has two extrema, yielding two AdS$_5$ solutions, located at
\begin{equation}
\sigma\,=\,0\qquad\qquad\qquad{\rm or}\qquad\qquad\qquad\cosh^2\sigma = \frac{3}{2}\,.
\end{equation}
The first preserves $\mathcal N=2$ supersymmetry and corresponds to the standard ${\rm AdS}_5\times\,$Sasaki--Einstein$_5$ vacuum of type IIB supergravity, while the second is non-supersymmetric and lifts to a solution originally found by Romans \cite{RomansIIBsols}.

It is clear that the RR scalar $C_0$ is a suitable axion verifying our general conditions to support a Lifshitz solution.\footnote{It may seem that $\chi$ could also be an appropriate axion because it appears in (\ref{5duniversal}) covered by a derivative. However, this is not the case since this scalar is charged already at the 5-dimensional level. As a consequence, after the circle reduction the equation of motion of the vector $A$ sets $D\chi = 0$ in our ansatz, meaning that the energy-momentum tensor of $\chi$ vanishes and therefore cannot support the Lifshitz metric. This example illustrates the necessity of the requirement made in section \ref{GeneralResult} that the axion is uncharged at the $(d+1)$-dimensional level. Note that one could alternatively solve the $A$ equation by taking $\sigma=0$, anyway this again gives a vanishing energy-momentum tensor for $\chi$.\label{FootnoteChi}} 
By performing a circle reduction, we obtain a 4-dimensional $\mathcal N=2$ theory containing gravity, one hypermultiplet and one vector multiplet. Moreover, we will put a flux on the circle so that the vector will turn out to be massive, providing, through the mechanism of section \ref{GeneralResult}, a Lifshitz solution for each of the two AdS$_5$ vacua. We stress that this will be a consistent truncation of type IIB supergravity on $S^1$ times any 5-dimensional squashed Sasaki--Einstein manifold to four dimensions. 

The ansatz for the 5-dimensional metric is (\ref{metric}), particularized to $d=4$.  
Moreover, we expand
\begin{equation}\label{ExpansionHatA}
\hat A=\,A\,+\,\alpha\,(d\vartheta+\mathcal A)\,,
\end{equation}
with $\alpha$ and $A$ a 4-dimensional scalar and a 1-form, respectively. 
From the scalar sector, as in the general discussion of section \ref{GeneralResult} we take all the scalars dependent just on the four spacetime coordinates $x$, except the axion, whose expansion is instead $C_0(x,\vartheta)  =C_0(x)+\,n\,\vartheta$. This means that we are introducing a flux through the circle of the RR 1-form field strength, $F_1^{\rm flux} = n\, d\vartheta$, so that  
\begin{equation}\label{F1Bianchi}
F_1\,=\,dC_0+\,F_1^{\rm flux} \,\equiv\, DC_0+ n\,(d\vartheta+\mathcal A)\,,
\end{equation}
where we have defined the covariant derivative $DC_0=dC_0-n\,\mathcal A$, inducing a St\"uckelberg coupling for the new vector $\mathcal A$ and thus providing a mass term for it. 

The reduction of the 5-dimensional action is then straightforward and gives
\begin{eqnarray}
S \!&=&\!\frac{1}{2\kappa_4^2}\,\int\,\bigg[\,\Big(\,R-\frac{3}{2}\,dT^2- \,\frac{3}{2}\,e^{-2T}\,d\alpha^2-\frac{1}{2}\,\cosh^2{\sigma}\,d\phi^2-\frac{1}{2}\,e^{2\phi}\,\cosh^4{\sigma}\,(DC_0)^2\nonumber\\[2mm]
&&\quad\quad\quad -\,2\,d\sigma^2- \frac{1}{2}\,\sinh^2{(2\sigma)}\,D\chi^2+\frac{1}{2}\,e^\phi\,\sinh^2{(2\sigma)}\,DC_0\lrcorner D\chi-\frac{1}{2}\,e^{3T}\,\mathcal F^2\nonumber\\[2mm]
&&\quad\quad\quad - \frac{3}{2}\,e^T\,\left(F+\alpha\,\mathcal F\right)^2-\, 2\, V\Big) *1+ \alpha^3 \mathcal F\wedge \mathcal F  + 3\alpha^2\,F\wedge \mathcal F + 3\alpha\,F\wedge F \,\bigg],\qquad
\label{eq:kineticterms}\end{eqnarray}
where $D\chi=d\chi-3\,A$ and the 4-dimensional potential $V$ reads
\begin{equation}\label{ScalPot4d}
V= \frac{3}{2}\,e^{-T}\cosh^2{\sigma}\big[\cosh{(2\sigma)}-5\big] + \frac{1}{4}n^2\,e^{-3T+2\phi}\cosh^2{\sigma}+\frac{1}{16}e^{-3T}\sinh^2{(2\sigma)}\left(n\,e^\phi+6\alpha\right)^2\!.
\end{equation}
It is immediate to check that this does not admit any extremum at finite value of the fields, so there are no AdS$_4$ solutions in the model.

%%%%%%%%%%%%%%%%%%%%%%%%%%%%%%%%%%%%%%%%%%%%%%%%%%%%%%%%%%%%%%%%%%
\subsection{Compatibility with $\mathcal N=2$ supergravity}\label{N=2UnivHyper4d}

The 4-dimensional action above is consistent with the structure of gauged $\mathcal N=2$ supergravity coupled to one vector multiplet and one hypermultiplet. 

To describe the couplings in the vector multiplet, we define the complex combination $z^1 = \alpha + i \,e^T$ and choose the symplectic holomorphic section $X^I \equiv (X^0,X^1)= (1,z^1)$.
We find that the suitable prepotential is
\begin{equation}
\mathcal F(X) \,=\, -\frac{(X^1)^3}{X^0}\,,
\end{equation}
corresponding to the special K\"ahler manifold $\frac{{\rm SU}(1,1)}{{\rm U}(1)}$.
Then the K\"ahler potential is
\begin{equation}
K  \,=\,  - \log \,(8\, e^{3T}) \,=\, -\log \left[\,i (z^1-\bar z^{1})^3\right] \,,
\end{equation}
yielding the K\"ahler metric $g_{1\bar 1} \,\equiv\, \partial_1\partial_{\bar 1}K \,=\, \frac{3}{4}\,e^{-2T}$, which matches the $\alpha$ and $T$ kinetic terms. The period matrix $\mathcal N_{IJ}$ following from (\ref{FormulaForN}) reads
\begin{equation}
{\rm Re}\,\mathcal N_{IJ} \,=\, \left(\begin{array}{cc} 
-2\alpha^3  &   3 \alpha^2\\ [2mm]
3 \alpha^2  &  -6 \alpha
\end{array}\right)\;,\qquad {\rm Im}\,\mathcal N_{IJ} \,=\, \left(\begin{array}{cc} 
-(e^{3T} + 3 \alpha^2 e^T)  & 3\alpha e^T  \\ [2mm]
 3\alpha e^T  &  -3e^T
\end{array}\right)\,.
\end{equation}
Under the identification $A^I \,=\, \tfrac{1}{\sqrt 2}(-\mathcal A, A)$, the vector kinetic and topological terms in (\ref{eq:kineticterms}) match the ones in the general action (\ref{4dN=2action}).

The remaining scalar kinetic terms in (\ref{eq:kineticterms}) define the sigma-model metric
\begin{eqnarray}
h_{uv}dq^udq^v &=&  d\sigma^2 +\tfrac{1}{4}\cosh^2{\sigma}\,d\phi^2 +\tfrac{1}{4}e^{2\phi}\cosh^4{\sigma}\,dC_0^2  \nonumber\\ [2mm] && +\, \tfrac{1}{4}\sinh^2{(2\sigma)}d\chi^2 - \tfrac{1}{4}e^\phi\sinh^2{(2\sigma)}\,dC_0\, d\chi\,,
\end{eqnarray}
where we identify $q^u=\{C_0,\phi,\sigma,\chi\}$. This is a quaternionic metric on the $\frac{{\rm SU}(2,1)}{{\rm U}(2)}$ coset manifold. Indeed, it can be mapped to the universal hypermultiplet metric given e.g. in \cite{StromingerUnivHyp}. Explicitly, this can be showed as follows. Introduce the 1-forms
\begin{eqnarray}
u &=& e^{-i\chi}\left[(\cosh\sigma)^{-1} d\sigma - \sinh\sigma(\tfrac{1}{2}d\phi + i\,d\chi) \right]\,,\nonumber\\ [2mm]
v &=& -\tanh\sigma\, d\sigma + i \sinh^2\sigma\, d\chi - \tfrac{1}{2} d\phi - \tfrac{i}{2}\,e^\phi\cosh^2\sigma\, dC_0\,,
\end{eqnarray}
which satisfy $h_{uv}dq^udq^v = u\otimes \bar u + v\otimes \bar v$.
The quaternionic vielbein $\mathcal U^{A\alpha}$ is then given by
\begin{equation}
\mathcal U^{A\alpha} = \frac{1}{\sqrt 2}\left(\begin{array}{cc} u & -\bar v \\ v & \bar u \end{array}\right)\,.
\end{equation}
We construct the SU(2) connection $\omega^x$, $x=1,2,3$, as
\begin{eqnarray}
\omega^1 &=& i(u-\bar u) \;=\; 2(\cosh\sigma)^{-1}\sin\chi \,d\sigma + \sinh\sigma\left( 2 \cos\chi\, d\chi - \sin\chi\, d\phi \right)\nonumber \\ [2mm] 
\omega^2 &=&  u + \bar u \;=\; 2(\cosh\sigma)^{-1}\cos\chi \,d\sigma - \sinh\sigma\left( 2 \sin\chi\, d\chi +\cos\chi\, d\phi \right) \nonumber \\ [2mm] 
\omega^3 &=& -\tfrac{i}{2}(v-\bar v) \;=\; -\tfrac{1}{2}\,e^\phi \cosh^2\sigma \,dC_0 + \sinh^2\sigma \, d\chi\,.
\end{eqnarray}
From the associated curvature 
$\Omega^x \equiv \Omega^x_{uv} dq^u\wedge dq^v$,
one can then define the triplet of almost complex structures $(J^x)^u{}_v := - h^{uw}\Omega^x_{wv}$, and verify that they satisfy the quaternionic ${\rm SU}(2)$ algebra
\be
J^xJ^y \,=\, -\delta^{xy}\mathbbone + \epsilon^{xyz} J^z,
\ee
proving that the manifold under consideration is quaternionic.

Comparing the covariant derivatives in our action with the general form (\ref{generalDq}), we see that the Killing vectors generating the isometries being gauged are
\begin{equation}
k_I \,=\, (k_0,k_1) \,=\,  (\,\sqrt 2\, n\,\tfrac{\partial}{\partial C_0}, \,\, -3\sqrt 2 \,\tfrac{\partial}{\partial \chi}\,)\,.
\end{equation}
For the associated Killing prepotentials $P^x_I$, solving the defining equation (\ref{DefPx}), we find
\begin{eqnarray}
P^{1}_I &=& \left(\,0,\;\,6\sqrt 2 \sinh\sigma\cos\chi\, \right)\,,\nonumber\\ [2mm]
P^2_I &=& \left(\,0,\;\, - 6\sqrt 2 \sinh\sigma\sin\chi \,\right)\,,\nonumber\\ [2mm]
P^{3}_I &=& \left(\, \tfrac{1}{\sqrt 2}\,n\,e^\phi \cosh^2\sigma,\;\;3\sqrt 2(\cosh^2\sigma -2) \,\right)\,.
\end{eqnarray}
Here, $k_0$ and $P^x_0$ encode the gauging by the graviphoton, generated in the circle reduction by the introduction of the $F_1$ flux term, while $k_1$ and $P^x_1$ are inherited from the 5-dimensional theory.
Using these data one can now evaluate the general formula (\ref{4dN=2ScalPot}) for the $\mathcal N=2$ scalar potential, and verify that it precisely matches the expression in (\ref{ScalPot4d}).

%%%%%%%%%%%%%%%%%%%%%%%%%%%%%%%%%%%%%%%%%%%%%%%%%%%%%%%%%%%%%%
\subsection{Lifshitz solutions, new and old}

From the general argument given in section \ref{GeneralResult} it follows that the 4-dimensional model will contain two Lifshitz solutions of the form (\ref{LifMetric}), in correspondence with the two AdS$_5$ vacua of the parent theory. Following our algorithm, the vector supporting this metric is taken of the form (\ref{AnsatzMathcalA}). We also assume that all scalars are constant and the second vector is parallel to the first $A=-\alpha\,\mathcal A$. In this way, the equations of motion for $A$ and $\alpha$ reduce to (\ref{VectorEquations}), which in this particular case reads 
\begin{equation}
\sinh^2(2\sigma)\,\left(n\,e^\phi+6\,\alpha\right)\,=\,0\,,
\end{equation}
and is solved by $\sigma=0$  or $\alpha = -\frac{1}{6}\,n \,e^\phi$.
Note that for vanishing $\sigma$ we are in the case in which (\ref{VectorEquations}) has non-maximal rank and therefore $\alpha$ remains as a modulus. We can then proceed with the rest of the equations and, as expected from the general analysis, the model under consideration contains precisely two Lifshitz solutions, with $\sigma$ being fixed at the extrema of the 5-dimensional potential. Both have 
\be
z=2\,,\qquad A = - \alpha\, \mathcal A \,,\qquad e^{-T} =  \lambda^{2/3}\,,
\ee
with arbitrary $\chi, \,C_0$, since they do not appear in the 5-dimensional potential. The first solution is then specified by
\be\label{Lifsol1}
\sigma=0\,,\qquad \alpha\;\;{\rm arbitrary}\,,\qquad L^2  = \lambda^{-2/3}\,,\qquad e^{2\phi}\lambda^{4/3}  = \frac{4}{n^2}\,,
\ee
while the second is
\be\label{Lifsol2}
\cosh^2\sigma = \frac{3}{2}\,,\qquad \alpha = -\frac{1}{6}\,n\,e^\phi\,,\qquad L^2  = \frac 89\lambda^{-2/3}\,,\qquad e^{2\phi}\lambda^{4/3}  = \frac{3}{4n^2}\,.
\ee
Due to the consistency of the truncation, these lift to solutions of type IIB supergravity. The first one was already described in \cite{DonosGauntlett_Lif} working directly in 10 dimensions; the metric on the 5-dimensional compact manifold is Sasaki--Einstein. The second provides a new embedding of Lifshitz geometries into string theory, associated with the same squashed Sasaki--Einstein metric which provides the Romans' AdS$_5$ solution.

Notice that having non-vanishing $F_1$ flux $n$ is crucial for the solution to exist. Besides the value of $n$, in the expressions above we have one additional free parameter: a natural choice can be to regard the value of the dilaton $\phi$ as free and fix the rest in terms of it. Another option is to keep $\lambda$ as arbitrary, in which case it would be convenient to choose $\lambda=1$, so that the `radius' $L$ of the Lifshitz$_4$ solution coincides with the one of the parent AdS$_5$ solution. Hence the string coupling constant is $e^\phi\sim 1/n$, and a large flux $n$ is needed in order to trust the supergravity approximation.

The 5-dimensional AdS solution at $\sigma =0$ has $\mathcal N=2$ supersymmetry,\footnote{A subtlety in this example is that, while the second supersymmetry condition (\ref{SusyAdSin4dLanguage}) for AdS$_5$ backgrounds is easily verified, one has $k^u_i \rho^i \neq 0$ and therefore the first condition seems violated. Actually this is not the case, since the full condition reads $\mathcal U_u^{A\alpha}k^u_i \rho^i = 0$, and in the coordinates used here the quaternionic vielbein is degenerate at $\sigma=0$, so it cannot be dropped. A reparameterization leading to an invertible $\mathcal U_u^{A\alpha}$ makes the Killing vector vanish at $\sigma=0$, thus fulfilling $k^u_i \rho^i = 0$.} so (\ref{Lifsol1}) provides an example of a Lifshitz solution preserving two supercharges. Since in this case the 5-dimensional compact manifold is Sasaki--Einstein, these are the solutions whose supersymmetry was studied from a 10-dimensional perspective in \cite{DonosGauntlett_Lif}.
Going through the general derivation of subsection \ref{AdStoLifSusyBreaking}, and recalling the various quantities in subsection \ref{N=2UnivHyper4d}, we find that for $\sigma =0$ and $n\neq 0$ the hyperino projector indeed gives $\vep^2 \,=\, \gamma^t \vep_1$.
Taking this into account, the constraint (\ref{EqforAlphaSusy}) from the gaugino equation requires
\be\label{Valuealpha}
\alpha \,=\, \frac{1}{6}\,n\,e^\phi\,.
\ee 
Eq. (\ref{GauginoEq4}) corresponds to the projectors
\be
\left(\gamma^r\gamma^t + \mathbbone\right)\vep_1 \,=\, 0\,,\qquad\qquad
\left(\gamma^r\gamma^t - \mathbbone\right)\vep_2 \,=\, 0\,,
\ee
while (\ref{IntegrabilityGauginoEq}) leads to $L^2\lambda^{2/3}=1$.
Note that the second projector follows from the first one together with the one from the hyperino equation. The gravitino equation is satisfied for $z=2$ as in the general proof above.
The last relation in (\ref{Lifsol1}) does not seem to follow from the supersymmetry conditions.

A couple of comments about relation (\ref{Valuealpha}) are in order. First, note that this was not fixed by the equations of motion, and is opposite to the one in the non-supersymmetric solution (\ref{Lifsol2}). Second, the fact that $\alpha$ cannot vanish means that the 5-dimensional vector $\hat A$, expanded as in (\ref{ExpansionHatA}), takes the pure gauge value $\hat A = \alpha\, d\vartheta$. The type IIB origin of this vector is in the 10-dimensional metric: it appears in a vielbein of the form $\eta + \hat A$, where the 1-form $\eta= (d\psi\, +$ connection on $B_{\rm KE}$) specifies the Sasaki--Einstein manifold as an $S^1$ fibration, with circle coordinate $\psi$, over a K\"ahler--Einstein base $B_{\rm KE}$. The two $S^1$'s parameterized by $\psi$ and $\vartheta$ together make a torus, with the modulus $T$ controlling the relative size of the circles and $\alpha$ describing their relative angle. We conclude that in the Lifshitz solution at hand the $S^1$ which is fibered over the K\"ahler--Einstein base is parameterized by $d\psi + \alpha\, d\theta$, and is therefore tilted with respect to the one of the AdS$_5\times\,$Sasaki--Einstein$_5$ solution.

%%%%%%%%%%%%%%%%%%%%%%%%%%%%%%%%%%%%%%%%%%%%%%%%%%%%%%%%%%%%%%
\section{A more involved example from $T^{1,1}$}\label{T11extension}

In the particular case in which the manifold admitting a Sasaki--Einstein structure is the $T^{1,1}$ coset space (also known as the base of the conifold), the $\mathcal N=2$ action in the previous section can be extended to a consistent truncation of type IIB supergravity providing a gauged $\mathcal N=2$ supergravity in 4 dimensions with two vector multiplets and three hypermultiplets. This is obtained by reducing on $S^1$ a 5-dimensional model obtained in \cite[sect.$\:$7]{T11}, and provides a supersymmetrization of the truncation given in \cite{DonosGauntlett_Lif}. We present it here since it has several new interesting features with respect to the truncation above. A main one is that the $T^{1,1}$ manifold has topology $S^2\times S^3$, thus it contains a 2-cycle, whose left-invariant representative in second cohomology was denoted $\Phi$ in \cite{T11}. Therefore in type IIB supergravity one can introduce NSNS and RR 3-form fluxes on $S^1\times S^2$, namely $H^{\rm flux}=n_2\,\Phi\wedge d\vartheta $ and $F_3^{\rm flux}=n_3\,\Phi\wedge d\vartheta$, with constant $n_2$, $n_3$. The Bianchi identities $dH=0$ and $dF_3= H\wedge F_1$ can be solved as
%\begin{eqnarray}\label{T11Bianchis}
%H&=&
%dB+\,H^{\rm flux}\,,\nonumber\\[2mm]
%F_3&=&
%dC_2-C_0\,H+n_1\,d\vartheta\wedge B+\,F_3^{\rm flux},\qquad
%\end{eqnarray}
\be\label{T11Bianchis}
H=
dB+\,H^{\rm flux}\,,\qquad\qquad
F_3=
dC_2-C_0\,H+n_1\,d\vartheta\wedge B+\,F_3^{\rm flux},\qquad
\ee
where the ansatz for $F_1$ is as in (\ref{F1Bianchi}), with flux parameter $n_1$. After the reduction, this will be reflected at the 4-dimensional level in the presence of three different would-be axions, denoted $\{C_0,\,b^\Phi,\,c^\Phi\}$. Depending on which flux $\{n_1,\,n_2,\,n_3\}$ is active in the reduction, the role of the charged axion will be played by either one of these fields.\footnote{As mentioned in \cite{T11}, NSNS and RR fluxes threading the 3-cycle of $T^{1,1}$ do not allow for AdS$_5$ solutions, therefore we do not include them here. This amounts to take $p=q=0$ in the equations there.} This yields a rich set of Lifshitz solutions to type IIB supergravity.

A related interesting feature is that the addition of the 3-form fluxes permits the stabilization of the relevant scalars in the potential, so that the reduced theory contains an AdS$_4$ solution. 
This vacuum, which to the best of our knowledge had not appeared before in the literature, is non-supersymmetric, though no signs of instability are found among the modes we kept.  

The degrees of freedom in the 5-dimensional model are:
\begin{eqnarray}
&&\{{\rm metric}, \hat A, \hat a_1, u+v\}\hspace{2.3cm}{\rm gravity+ 1\; vector\; multiplet}\nonumber\\[2mm]
&&\{b^\Phi, b^\Omega, c^\Phi, c^\Omega, a, u, \phi, C_0, t, \zeta\}\hspace{2.5cm}{\rm 3 \;\; hypermultiplets.}
\end{eqnarray}
The higher-dimensional origin of these fields and the notation are explained in \cite{T11}, where the complete 5-dimensional action is also given.\footnote{Here we name $\zeta$ the scalar that was called $\theta$ there, in order to avoid confusion with the spacetime vielbeine or the circle coordinate.} It contains the same AdS$_5$ vacua as the truncation to the universal hypermultiplet of section \ref{ExampleSE}, which indeed can be retrieved from this one by switching off $\{b^\Phi, \,c^\Phi, \,a,\, t,\, \zeta\}$ and identifying (for $k=2$)
\be
-u=v=\frac12\,\log{(\cosh{\sigma})}\,,\quad\qquad c^\Omega=b^\Omega\,\tau=e^{\phi/2}\,e^{i\chi}\,\tau\,\tanh{\sigma}\,,\qquad\quad\hat a_1=-\hat A\,,
\ee
where $\tau=C_0+i\,e^{-\phi}$ is the axio-dilaton of type IIB supergravity.

Starting from this 5-dimensional theory, we perform a circle reduction as detailed above and go to 4 dimensions. With respect to the example in the previous section, here we have a richer set of possibilities for the identification of the axionic symmetry that is crucial to find Lifshitz solutions. A suitable expansion ansatz for the 5-dimensional scalars allowing to consider these options all together, and corresponding to the 10-dimensional ansatz (\ref{T11Bianchis}), is
\begin{eqnarray}
C_0(x,\vartheta)&=&C_0(x)+\,n_1\,\vartheta\nonumber\\[2mm]
b^\Phi(x,\vartheta)&=&b^\Phi(x)+\,n_2\,\vartheta\nonumber\\[2mm]
c^\Phi(x,\vartheta)&=&c^\Phi(x)+\,n_3\,\vartheta+\,n_1\,b^\Phi(x)\,\vartheta+\frac{1}{2}\,n_1\,n_2\,\vartheta^2\\[2mm]
c^\Omega(x,\vartheta)&=&c^\Omega(x)+\,n_1\,b^\Omega(x)\,\vartheta\nonumber\\[2mm]
a(x,\vartheta)&=&a(x)+\frac12\,\vartheta\,\left[n_2\,c^\Phi(x)-n_3\,b^\Phi(x)\right]+\frac14\,n_1\,n_2\,b^\Phi(x)\,\vartheta^2+\frac{1}{12}\,n_1\,n_2^2\,\vartheta^3.\nonumber
\end{eqnarray}
Here, the $\vartheta$-dependence is chosen in such a way that plugging the ansatz into the 5-dimensional field strengths given in \cite{T11}, the explicit dependence on the $S^1$ coordinate $\vartheta$ drops out. In this way the circle reduction is possible even if the expressions above are not U(1) singlets. The remaining scalars are assumed to depend just on the 4-dimensional spacetime coordinates, while the vectors are expanded as usual:
\begin{equation}
\hat A=A+\,\alpha\,(d\vartheta+\mathcal A)\,,\qquad\qquad\qquad\quad\hat a_1=a_1+\,\beta\,(d\vartheta+\mathcal A)\,,
\end{equation}
where $\alpha$ and $\beta$ are scalars in four dimensions.
The reduction on the circle proceeds smoothly and leads to the following 4-dimensional action
\begin{equation}
S\;=\;\frac{1}{2\kappa_4^2}\int \left( R -2V \right) *1 \,\,+\, S_{\rm vect}\,+\,S_{\rm kin,scal}  \,,
\end{equation}
where the kinetic and topological terms of the gauge fields are
\begin{eqnarray}\label{VectActionT11}
 S_{\rm vect}\!\!\!&=&\!\!\!\frac{1}{2\kappa_4^2}\int\bigg\{\Big[-\frac12e^{3T}\left(d\mathcal A\right)^2 -\frac12 e^{\frac83u+\frac83v+T}(dA + \alpha\,d\mathcal A)^2 -e^{-\frac{4}{3}u-\frac{4}{3}v+T}(da_1+\beta\,d\mathcal A)^2\Big]*1\nonumber\\ [2mm]
&& \quad\qquad+\,\alpha\,(da_1+\beta\,d\mathcal A)\wedge(da_1+\beta\,d\mathcal A)+2\,\beta\,dA \wedge da_1+\,\beta^2\,dA \wedge d\mathcal A\bigg\}\,,
 \end{eqnarray}
while for the kinetic terms of the scalars we have
\begin{eqnarray}
 S_{\rm kin,scal}\!\!\!&=&\!\!\! -\frac{1}{2\kappa_4^2}\int\bigg\{\,\frac32 dT^2 + \frac{28}{3}du^2 + \frac{4}{3}dv^2 + \frac{8}{3}du\lrcorner dv + \frac{1}{2}d\phi^2 + \frac{1}{2}e^{2\phi}(DC_0)^2 \nonumber\\[2mm]
 &&\qquad + \,dt^2 + \sinh^2 t \,D\zeta^2 + \frac{1}{2}e^{\frac{8}{3}u+\frac{8}{3}v-2T}\,d\alpha^2 + e^{-\frac{4}{3}u-\frac{4}{3}v-2T}\,d\beta^2  \, \nonumber\\[2mm]
&&\qquad   +\, e^{-4u-\phi}\Big[\cosh(2t)\,(h_1^\Phi)^2+\,\cosh^2 t \,|h_1^\Omega|^2- \sinh^2 t\, {\rm Re}\left(e^{-2i\zeta} (h_1^\Omega)^2\right)\nonumber\\[3mm]
&& \qquad  +\,2\sinh(2t)\,h_1^\Phi\lrcorner\, {\rm Re}\big(i \,e^{-i\zeta} h_1^\Omega \big) \Big] + e^{-4u+\phi}\Big[h\,\rightarrow\,g\Big] + 2\,e^{-8u}(f_1)^2\bigg\} *1,\qquad\;\; 
\end{eqnarray}
with the charged field $D\zeta \,=\, d\zeta -3 A\,$. Finally, the potential takes the cumbersome form
\begin{eqnarray}\label{potential4d}
 2\,V\!\!&=&\!-\,e^{-\frac83u-\frac23v-T}\Big[24\,e^{-2u}\cosh t-9\,e^{-2v}\sinh^2t-4\,e^{-4u+2v}\Big]\nonumber\\[2mm]
 &&+\,2\,e^{-\frac{32}{3}u-\frac{8}{3}v-T}f_0^2  +  2\,e^{-8u-3T}\,j_0^2 + \frac{n_1^2}{2}\,e^{2\phi-3T} + 9\,e^{-3T}\,\alpha^2\sinh^2t\nonumber\\[2mm]
&& +\, e^{-\frac{20}{3}u-\frac{8}{3}v-T-\phi}\Big[\cosh^2t\,|h_0^\Omega|^2-\sinh^2t\,\,{\rm Re}\left(e^{-2i\zeta}(h_0^\Omega)^2\right)\Big] \nonumber\\[2mm]
&& +\, e^{-\frac{20}{3}u-\frac{8}{3}v-T+\phi}\Big[\cosh^2t\,|g_0^\Omega|^2-\sinh^2t\,\,{\rm Re}\left(e^{-2i\zeta}(g_0^\Omega)^2\right)\Big] \nonumber\\[2mm]
&& +\, e^{-4u-3T-\phi}\Big[\,n_2^2\cosh{(2t)}+\alpha^2\cosh^2t\,|h_0^\Omega|^2-\alpha^2\sinh^2t\,\,{\rm Re}\left(e^{-2i\zeta}(h_0^\Omega)^2\right)\nonumber\\[2mm]
&&+\,2\,n_2\,\alpha\sinh{(2t)}\,{\rm Im}\left(e^{-i\zeta}h_0^\Omega\right)\Big] + e^{-4u-3T+\phi}\Big[\,\cosh{(2t)}(n_1\,b^\Phi+n_3-n_2\,C_0)^2\nonumber\\[2mm]
&&+\,\cosh^2t\,|\alpha\,g_0^\Omega-n_1\,b^\Omega|^2-\,\sinh^2t\,\,{\rm Re}\left(e^{-2i\zeta}(\alpha\,g_0^\Omega-n_1\,b^\Omega)^2\right)\nonumber\\[2mm]
&&+\,2\,\sinh{(2t)}\,(n_1\,b^\Phi+n_3-n_2\,C_0)\,\,{\rm Im}\left(\,e^{-i\zeta}(\alpha\,g_0^\Omega-n_1\,b^\Omega)\right)\Big] \,.
\end{eqnarray}
In these expressions, we have the following identifications, coming from the expansion of the NSNS and RR field strengths of type IIB supergravity:
\begin{equation}
\begin{array}{rclcrcl}
h_1^\Omega&=&(d-3iA)b^\Omega\equiv Db^\Omega\,,&\qquad\qquad\qquad\qquad&g_1^\Omega&=&Dc^\Omega-C_0\,Db^\Omega\,,\nonumber\\[2mm]
h_0^\Omega&=&3ib^\Omega\,,&\qquad\qquad\qquad&g_0^\Omega&=&3i(c^\Omega-C_0\,b^\Omega)\,,\nonumber\\[2mm]
h_1^\Phi&=&db^\Phi-n_2\,\mathcal A\equiv Db^\Phi\,,&\qquad\qquad\qquad\qquad& g_1^\Phi&=&Dc^\Phi-C_0\,Db^\Phi\,,
\end{array}
\end{equation}
and
\begin{eqnarray}
f_0&=&k+3\,{\rm Im}\big[b^\Omega\,\overline{c^\Omega}\big]\nonumber\\[2mm]
j_0&=&-2\beta-\alpha\,f_0+\frac{n_1}{2}\Big[|b^\Omega|^2-(b^\Phi)^2\Big]+n_2\,c^\Phi-n_3\,b^\Phi \label{Expansf} \\[2mm]
f_1&=&Da + \frac12\left[{\rm Re}(b^\Omega\overline{Dc^\Omega})-b^\Phi D c^\Phi  - {\rm Re}(c^\Omega\overline{Db^\Omega}) + c^\Phi D b^\Phi\right],\nonumber
\end{eqnarray}
where the combination $j_0$ comes from the expansion of $\hat f_1=f_1+j_0\,(d\vartheta+\mathcal A)$. The constant $k$ appearing in $f_0$ parameterizes the RR 5-form flux.
The yet unspecified covariant derivatives are 
\begin{eqnarray}
Dc^\Omega &=& dc^\Omega - 3iA\, c^\Omega -n_1\,b^\Omega\mathcal A\,,\qquad\qquad\qquad\qquad Dc^\Phi = dc^\Phi-n_3\,\mathcal A -n_1\,b^\Phi\mathcal A\,,\nonumber\\[2mm]
Da&=&da-2a_1-k\,A +\frac{1}{2}(n_3\,b^\Phi-n_2\,c^\Phi)\mathcal A\,.
\end{eqnarray}

We checked that the action matches the structure of gauged $\mathcal N=2$ supergravity coupled to two vector multiplets and three hypermultiplets. Regarding the vector multiplets, we define the complex scalars
\begin{eqnarray}
\frac{X^1}{X^0}\,=\,z^1 \,=\, \alpha + i \,e^{-\frac{4}{3}(u +v) + T},\qquad\qquad \frac{X^2}{X^0}\,=\,z^2\,=\, \beta + i\,e^{\frac{2}{3}(u + v) + T}\,.
\end{eqnarray}
and find that the suitable prepotential is 
\begin{equation}
\mathcal F(X) \,=\, -\frac{X^1 (X^2)^2}{X^0}\,,
\end{equation}
associated with the special K\"ahler manifold $\left(\frac{{\rm SU}(1,1)}{{\rm U}(1)}\right)^2$.
Then the K\"ahler potential is
\begin{equation}
K \,=\, -\log \left( 8\, e^{3T}\right) \,=\, -\log \left[i\,(z^1-\bar z^1)(z^2-\bar z^2)^2\right],
\end{equation}
whose K\"ahler metric $g_{i\bar \jmath} = \partial_i\partial_{\bar \jmath}K$ reproduces the kinetic terms of the scalars $\alpha,\, \beta,\, T$ and $(u+v)$.
Using (\ref{FormulaForN}), one can compute the period matrix $\mathcal N_{IJ}$ and, under the identification $A^I \,=\, \tfrac{1}{\sqrt 2}(-\mathcal A,\, A,\, a_1)$, precisely match the vector terms (\ref{VectActionT11}) with the ones in the general $\mathcal N=2$ action (\ref{4dN=2action}).
Though we will not report the details here, we also checked that the remaining scalars parameterize the quaternionic manifold $\frac{{\rm SO}(4,3)}{{\rm SO}(4)\times {\rm SO}(3)}\,$. Finally, identifying the Killing vectors which generate the gauged isometries, computing the associated Killing prepotentials, and evaluating the general formula (\ref{4dN=2ScalPot}), we verified that the scalar potential (\ref{potential4d}) is recovered. 

We now discuss the Lifshitz solutions of this model. The main feature we want to emphasize is that, as already mentioned above, the role of the axion can be played by either $C_0$, $b^\Phi$ or $c^\Phi$ depending if $n_1$, $n_2$ or $n_3$ respectively are active.\footnote{Parallel arguments to the ones given in footnote \ref{FootnoteChi} exclude the use of the scalar $a$ as a suitable axion.}
Since the 5-dimensional model admits the same two AdS solutions considered in the previous section, the 4-dimensional action contains two Lifshitz solutions with $z=2$, which are easily found following the procedure in section \ref{GeneralResult}. These lift to type IIB solutions, generically with all NSNS and RR fluxes turned on \cite{DonosGauntlett_Lif}. The equation for $\mathcal A$ reads in this case
\begin{equation}
\frac{2\,z}{L^2} \,=\, \lambda^2\,\Big[\,n_1^2\,e^{2\phi}+2\,n_2^2\,e^{-\phi}+2\,(n_1\,b^\Phi+n_3-n_2\,C_0)^2\,e^\phi\Big],
\end{equation}
and we see that the solution can be supported by any of the three fluxes $\{n_1\,,n_2\,,n_3\}$. Let us stress that for the supersymmetric solution the equations for the vectors $A$ and $a_1$ as well as the scalars $\alpha$ and $\beta$ are all solved by taking $j_0=0$, which fixes the value of $\beta$, leaving $\alpha$ unconstrained. The supersymmetry equations can be solved as in subsection \ref{AdStoLifSusyBreaking}, with condition (\ref{EqforAlphaSusy}) fixing $\alpha$.

Interestingly, the model also contains and AdS$_4$ solution, as one can check by extremizing the potential (\ref{potential4d}). This is located at
\begin{equation}\nn
(n_1\,b^\Phi -n_2\,C_0 + n_3)\,=\,(2\beta + k\, \alpha + \frac{1}{2}\,n_1 (b^{\Phi})^2 + n_3\, b^{\Phi} - n_2\, c^{\Phi}) \,=\,b^\Omega\,=\,c^\Omega\,=t\,=\,0
\end{equation}
and
\be
e^{8u}=\frac{16}{45}k^2\,,\qquad e^{8v}=\frac{9}{80}k^2\,,\qquad e^{6T}=\frac{4}{405}k^2 n_1^2\,n_2^4\,,\qquad e^{6\phi}= \frac{45}{16}\frac{n_2^4}{k^2n_1^4}\,. 
\ee
The cosmological constant, given by the value of the potential at the extremum, is 
\be
\Lambda \,=\, -\frac{27}{8}\frac{ 3^{2/3}\,5^{5/6}}{(2\,k^5\,n_1\, n_2^2)^{1/3}}\,.
\ee
In addition to $\zeta$ and $a$, one has three other moduli chosen from $\{b^\Phi,\,c^\Phi,\,C_0,\,\alpha,\,\beta\}$. Again, the consistency of the truncation guarantees the lifting of the solution to type IIB supergravity. Generically all the NSNS and RR fluxes are switched on. While the RR 3-form flux $n_3$ can be zero, having non-vanishing RR 1-form flux $n_1$, RR 5-form flux $k$, and NSNS 3-form flux $n_2$ is crucial to stabilize the metric moduli $u,\,v\,,T$ and the dilaton $\phi$. The solution does not appear to preserve any supersymmetry. We checked that for the modes we kept the spectrum of fluctuations around the vacuum is non-negative. However we cannot claim stability within the full type IIB supergravity, since we have no access to the modes out of the truncation that can develop large negative masses.

The simultaneous presence of Lifshitz$_4$ and AdS$_4$ solutions is a prominent characteristic of our model. An interpolation between these spacetimes was considered at the phenomenological level already in \cite{Kachru} and describes, in the dual theory, restoration of conformal invariance at the endpoint of the renormalization group flow. The model provided here is suitable to embed this flow into string theory.

%%%%%%%%%%%%%%%%%%%%%%%%%%%%%%%%%%%%%%%%%%%%%%%%%%%%%%%%%%%%%%%
\section{Further examples and prospects}\label{Conclusion}

The general results of this paper can be applied to other instances than the ones in sections \ref{ExampleSE} and \ref{T11extension}, most interestingly to other known reductions of string theory to diverse dimensions meeting the requirements in section \ref{GeneralResult}. For example, one can reduce type IIB supergravity on a 5-dimensional Einstein manifold with a Freund--Rubin ansatz for the RR 5-form and keeping the axio-dilaton. In this way one obtains a consistent truncation to 5-dimensional Einstein gravity plus the axio-dilaton and a cosmological constant. Essentially, this is the action (\ref{5duniversal}) with $\hat A=\sigma=0$, which indeed is a consistent truncation. This model contains an AdS$_5$ vacuum, so we can use the axion to support the flux on the circle and get a simple description of the general ${\rm Lifshitz}_4$ solutions of type IIB supergravity based on internal Einstein manifolds \cite{DonosGauntlett_Lif}. T-duality along the circle should lead to a simple Lifshitz$_4$ solution of massive type IIA supergravity.

We have also checked the existence of Lifshitz$_4$ solutions starting from the more involved 5-dimensional gauged $\mathcal N=4$ supergravity which arises as a consistent truncation of type IIB on the $T^{1,1}$ coset preserving the full set of ${\rm SU(2)}\times {\rm SU(2)}$ invariant modes \cite{T11, Bena}. As found in \cite{T11}, this model contains an entire family of AdS$_5$ vacua, from which one is able to construct correspondingly a family of Lifshitz$_4$ solutions with non-Einstein internal metrics (thus not included in the results of \cite{DonosGauntlett_Lif}). An important point is that this $\mathcal N=4$ truncation contains both 2-forms and non-Abelian gaugings, which were not considered in the algorithm of section~\ref{GeneralResult}. Nonetheless the Lifshitz solutions are still present for an ansatz with vanishing 2-forms, hinting to a possible extension of the proof in section \ref{GeneralResult} to include forms of higher degree and more complicated interactions.

If such a generalization is viable, it will be interesting to apply it to the known AdS$_5$ vacua of 5-dimensional $\mathcal N=8$ supergravity with SO(6) gauge group, which is believed to represent a consistent truncation of type IIB on $S^5$. In particular, one could see if for the round sphere, supporting maximally supersymmetric AdS$_5$ solutions, the Lifshitz descendants can also preserve more than the just two supercharges found in our analysis. Indeed this is what happens in the case of non-relativistic solutions with Schr\"odinger invariance \cite{Bobev:2009mw,Donos:2009zf}.
A prominent vacuum of $\mathcal N=8$ supergravity is the $\mathcal N=2$ Pilch--Warner solution \cite{KhavaevPilchWarner, PilchWarner}, which can anyway be obtained from an $\mathcal N=2$ subtruncation of the $\mathcal N=8$ theory \cite{Pilch:2000fu, ChargedClouds}, and therefore falls into the class of actions considered in this work. 
We thus obtain a new supersymmetric Lifshitz$_4$ solution of type IIB supergravity.
The truncation contains both the round sphere and the Pilch--Warner solutions, as well as the radial flow interpolating between the two. A natural question is whether there exists a similar flow connecting the corresponding anisotropic solutions.

Another possible interpolating solution that could be studied within our setup is the one connecting  AdS$_{d+1}$ (with a compact direction) with the $(d+1)$-dimensional lift of the Lifshitz$_d$ solution, corresponding to a geometry with Schr\"odinger invariance and dynamical exponent $z=0$. Progresses towards a holographic interpretation of this deformation have been made in \cite{Costa}.

Finally, we stress that the class of truncations discussed in this paper is also suitable for studying other types of solutions, for instance black holes with Lifshitz asymptotics. Since charged scalars are present, our models should also be applicable to the holographic study of superconductors, in particular to the question of their ground state, argued to be of Lifshitz type in \cite{GubserNellore}.

%%%%%%%%%%%%%%%%%%%%%%%%%%%%%%%%%%%%%%%%%%%%%%%%%%%%%%%%%%%%%%%
\bigskip
\section*{Acknowledgments}

\noindent We are particularly grateful to Gianguido Dall'Agata for many insightful discussions and for comments on the manuscript. We also thank Jerome Gauntlett, Monica Guica, Nick Halmagyi, Jelle Hartong, Paul Koerber and Michela Petrini for stimulating conversations.
DC is supported by the Fondazione Cariparo Excellence Grant {\em String-derived supergravities with branes and fluxes and their phenomenological implications}.

%%%%%%%%%%%%%%%%%%%%%%%%%%%%%%%%%%%%%%%%%%%%%%%%%%%%%%%%%%%%%%%
\appendix

\section{4-dimensional $\mathcal N=2$ supergravity conventions}\label{conventions}

Contrary to \cite{N=2review}, we work with a $(-+++)$ spacetime signature. Then $\,(\gamma^\mu)^{\rm here}= i(\gamma^\mu)^{\rm there}$.
Another difference with respect to \cite{N=2review} is that our differential forms are defined including the combinatorial weight. In particular, for the gauge field strengths we have $F^I= dA^I= \frac{1}{2}F^I_{\mu\nu}dx^\mu\wedge dx^\nu$, while in the conventions of \cite{N=2review} $F^\Lambda= dA^\Lambda= \mathcal F^\Lambda_{\mu\nu}dx^\mu\wedge dx^\nu$, hence $(F^I_{\mu\nu})^{\rm here}=2(\mathcal F^\Lambda_{\mu\nu})^{\rm there}$.
For the Levi-Civita symbol we take $\epsilon_{0123}=+1$.

As in \cite{N=2review}, the Sp(1)$\,\cong\,$SU(2) metric $\epsilon_{AB}$ satisfies $\epsilon_{AB}=-\epsilon_{BA}$ and $\epsilon_{AB}\epsilon^{BC}= -\delta_A^C$, while $(\sigma_x)_A{}^B$ are the standard Pauli matrices. On bosonic quantities, the SU(2) indices are raised and lowered according to the SW-NE convention, namely $\epsilon_{AB}V^B=V_A$ and $V_B\epsilon^{BA}=V^A$. On the fermions, the SU(2) indices are located according to chirality; for instance, $\vep_A$ has positive chirality, while $\vep^A$ has negative chirality.

We choose our Clifford algebra conventions in such a way that the 4-dimensional gamma matrices are all real, and charge conjugation coincides with complex conjugation. Then $\gamma_5:=\,i\,\gamma^t\gamma^1\gamma^2\gamma^r$ is purely imaginary, so we have that complex conjugation flips the chirality. Hence we can take $\vep^A \equiv (\vep_A)^*$.

We are also fixing the gauge coupling constant $g=1$.

%%%%%%%%%%%%%%%%%%%%%%%%%%%%%%%%%%%%%%%%%%%%%%%%%%%%%%%%%%%%%%

\end{document}